\documentclass[11pt,a4article]{article}

\usepackage{jheppub}
\usepackage[T1]{fontenc} 
\usepackage{epsfig,epsf}
\usepackage{amsmath}
\usepackage{amsthm}
\usepackage{amsfonts}
\usepackage{amssymb}
\usepackage{caption}
\usepackage{dsfont}
\usepackage{mathrsfs}
\usepackage{epstopdf}
\usepackage{multirow}

\usepackage{comment}

\usepackage{rotating}

\usepackage{marvosym}

\usepackage{soul}
\usepackage{xcolor}

\usepackage{caption}
\usepackage{subcaption}
\usepackage{array}   
\newcolumntype{C}{>{$}c<{$}}

\usepackage{slashed}

\def\II{\hbox{{1}\kern-.25em\hbox{l}}}

\DeclareMathOperator{\Li}{Li}

\def\II{\hbox{{1}\kern-.25em\hbox{l}}}

\newtheorem{proposition}{Proposition}

\newcommand{\iu}{{\sf i}}
\renewcommand{\imath}{\mathrm{i}}
\renewcommand{\Re}{\operatorname{Re}}
\renewcommand{\Im}{\operatorname{Im}}
\renewcommand{\phi}{\varphi}

\def\II{\hbox{{1}\kern-.25em\hbox{l}}}

\def\lb{\label}
\def\be{\begin{equation}}
\def\ee{\end{equation}}

 \usepackage{tocloft}

\numberwithin{equation}{section}

\title{
{\textsc
Conformal four-point
ladder integrals in diverse dimensions
   and polylogarithms}
}

\author[a, b, f]{S. E. Derkachov,}
\author[c, d]{A. P. Isaev}
\author[e]{and L. A. Shumilov}

\affiliation[a]{St.Petersburg Department of the Steklov Mathematical Institute of Russian Academy of Sciences,Fontanka 27, 191023, St.Petersburg, Russia}

\affiliation[b]{Saint Petersburg State University, 7/9 Universitetskaya Embankment, Saint Petersburg, Russia}

\affiliation[c]{Laboratory of Theoretical Physics,
Joint Institute for Nuclear Research,
6 Joliot-Curie Str, 141980, Dubna,
Moscow Region, Russia}

\affiliation[d]{Lomonosov Moscow State University, Physics Faculty, Russia}

\affiliation[e]{
   II. Institut f\"ur Theoretische Physik, Universit\"at Hamburg, Luruper Chaussee 149,
   D-22761 Hamburg, Germany}

\affiliation[e]{Beijing Institute of Mathematical Sciences and Applications,
		Huairou district, Beijing, 101408, China }

\emailAdd{derkach@pdmi.ras.ru}

\emailAdd{isaevap@theor.jinr.ru}

\emailAdd{leonid.shumilov@desy.de}

\abstract{
In the paper, a family of conformal four-point ladder diagrams in arbitrary space-time dimensions is considered. We use the representation obtained via explicit calculation using the operator approach and conformal quantum mechanics to study their properties such as symmetries, loop and dimensional shift identities. In even dimensions, latter allows one to reduce the problem to the two-dimensional case, where notable factorization holds. Additionally, for a specific choice of propagator powers, we show that the
 representation can be written
in the form of linear combinations of
classical polylogarithms (with coefficients that are rational functions) and explore the structure of the resulting expressions.
}


\begin{document}
\maketitle

\section{Introduction}










In recent years, significant progress has been made in the computation of scattering amplitudes in various quantum field theories (see e. g.~\cite{BDKo1, BDKo2, ElvH, BDKo3, BDKo4, arkh}). In calculation of physical observables, e.g. in QCD, the algoritmizability and uniformity are often prioritized over the preserving specific symmetries at intermediate steps. As a result, techniques such as integration-by-parts 
 (IBP)~\cite{Chetyrkin:1981qh}
  or differential equation methods~\cite{Henn:2013pwa, Rem, Lee},
   have undergone significant development. On the other hand, in theories with enhanced symmetries, it is sometimes possible to exploit these symmetries directly in computations. A well-known example of such a theory is $\mathcal{N} = 4$ SYM, especially in the planar limit. The underlying superconformal symmetry made it possible to fully constrain tree-level MHV amplitudes in a compact form~\cite{Parke:1986gb, Nair:1988bq} even in the early years of study.

\vspace{0.2cm}
Remarkably, the idea of using specific symmetries in Feynman integrals was implemented even before the establishment of IBP-reduction (which is nowadays widely used as a universal tool), leading to an equally universal approach. Gegenbauer polynomial technique~\cite{ChKT, Kot1} (see also~\cite{Kot2} for a modern review) deserves special mention. It utilizes the $D$-dimensional rotational symmetry, which is the signature of all Feynman integrals. Despite challenges in its algorithmic implementation, this technique provided a possibility to obtain a series of important results~\cite{Celmaster:1980ji, Terrano:1980af, Lampe:1982av} and continues to reveal deep connections with other methods~\cite{Derkachev:2022lay}.

\vspace{0.2cm}

However, not all symmetries of the theory can be read directly from its Lagrangian formulation. A notable example is the conjectured $\text{AdS}/\text{CFT}$ correspondence, which relates planar $\mathcal{N} = 4$ SYM to string theory on $\text{AdS}_5\times S^5$. This duality in turn reveals the underlying integrable structure. For instance, in the so-called fishnet theory, which can be obtained as a deformation of planar $\mathcal{N} = 4$ SYM~\cite{GK,GGKK,KO, Caetano:2016ydc, KaSt}, underlying integrability leads to significant results in the calculation of
four-point Basso-Dixon correlators~\cite{BD,BD1,DKO,DO1,DO2,DO3,DFO,Duhr1,Duhr2} and more general correlators~\cite{GKK,GKKNS,CK,BCF,O1,O2,AO}. Another hidden symmetry, crucial to the structure of planar $\mathcal{N} = 4$ SYM, is the dual conformal symmetry~\cite{AR, DKS, DHKS}, which together with ordinary 
conformal
symmetry forms a Yangian 
symmetry~\cite{DHP}. This structure not only constrains tree-level amplitudes but can also be extended to the loop level, as reflected in the BDS ansatz for all-loop MHV amplitudes~\cite{Bern:2005iz}. Note that dual conformal symmetry also finds its application beyond the four-dimensional theories~\cite{Kaz1, BIKT, BL, BKKTV}.

\vspace{0.2cm}

From a technical point of view, dual conformal symmetry manifests itself as conformal invariance of the integrals associated with dual Feynman graphs. These integrals have attracted attention since the early days of multiloop calculations and continue to be actively studied~\cite{GorIs, DHSS}. Conformal symmetry at the level of
perturbative integrals makes it possible to use powerful analytical identities, such as e.g. the star–triangle relation~\cite{Zam, Vasiliev:1981dg, Chicherin:2012yn} or the theory of graphical functions~\cite{Brown:2012ia, Schnetz:2013hqa}, to simplify calculations dramatically.
The star–triangle relation (interpreted
as the Yang-Baxter equation; for a brief review see the
subsection 5.3 in
\cite{Isaev:2022mrc}) allows one to consider not only individual diagrams but entire families, such as fishnet diagrams \cite{Zam} or conformal
ladder diagrams in arbitrary dimensions $D$
\cite{Isa, Isaev:2007uy}. Recall that
 ladder diagrams were first studied by Ussyukina and Davydychev, who derived all-loop results in four dimensions~\cite{Usyukina:1992jd, Usyukina:1993ch}. Later, a generalization of conformal ladder integrals were analyzed, and their analytic properties, such as single-valuedness, were explored in~\cite{Drum}.

\vspace{0.2cm}

Note that in the paper \cite{Chicherin:2012yn}
we found the general $D$-dimensional
conformal invariant solution to the Yang-Baxter equation
that generalizes the $1$-dimensional solution of
\cite{Isaev:2007uy} and the $2$-dimensional solution of
\cite{DKM} and underlies
Lipatov's integrable models describing high-energy
behaviour of QCD as well as
integrable structures of scattering amplitudes in $\mathcal{N}=4$ SUSY
\cite{Lip1,Lip2,FaKo}.
This Yang-Baxter solution was used,
for example, in the
investigations of the Yangian symmetries of perturbative
integral in fish-net type, loom and checkerboard CFT
\cite{Kazakov:2023nyu, Kazakov:2022dbd, Alfim}.

\vspace{0.2cm}

In one of our 
previous
works~\cite{DIS2}, we used the graph-building operator technique~\cite{GK,GGKK,KO,DKO, Gromov:2018hut, DO1, DO2, DIS1} together with the connection to conformal quantum mechanics~\cite{Isa, Isaev:2007uy} to obtain an all-loop result for the conformal ladder and zig-zag four-point correlators in arbitrary dimensions. While our expression for ladder diagrams
was fully analytical and valid in any dimension, it was formulated in terms of Gegenbauer polynomials, so despite its generality, the form of the answer may be of little practical use. Recently, remarkable progress has been made in two dimensions, where conformal ladder integrals were shown to be related to twisted partition functions~\cite{Petkou0,Petkou, Petkou2}, allowing them to be rewritten in terms of classical polylogarithms. In the present work, we show that in arbitrary even dimensions and a specific choice of propagator indices our previous representation can be systematically expressed using classical polylogarithms and rational functions. Furthermore, we verify that our representation satisfies a loop and dimensional shift identity studied in~\cite{Loebb, Petkou, Petkou2}. The latter, among other things, allows one to express the answer for $D$-dimensional diagram in terms of two-dimensional, where notable factorization holds (see~\cite{DKO, Derkachev:2022lay} for examples of such a factorization). We believe that these results can be useful not only from a practical point of view but also in revealing possible underlying symmetries, such as antipodal self-duality~\cite{Dixon:2025zwj} in the fishnet theory which also holds for the one-loop ladder diagrams in $D = 4$. Before closing this section, we would like to mention that despite the rich diversity of results the studies of conformal integrals continue to attract interest~\cite{Alkalaev:2025fgn, Alka, He:2025lzd}.
\vspace{0.2cm}

The paper is organized as follows. In Section~\ref{sect:2} we review the calculation of a general family of conformal four-point ladder diagrams using the graph-building operator and conformal quantum mechanics. After discussing the most general choice of parameters for ladder diagrams, we restrict ourselves to the case described by a single parameter $\beta$. We show that the obtained representation admits a dimensional shift $D \mapsto D + 2$ for a general choice of $\beta$. Additionally, we show that the combination of dimensional shift operator with the graph-building operator allows us to construct an operator that increases the number of loops $L \mapsto L + 1$. In Section~\ref{sect:3} we move to the case of ladder diagrams with the specific parameter $\beta = 1$ and present the derivation of the answer containing only classical polylogarithms and rational functions. Then, we study a representation based on a two-dimensional factorized form in the case $\beta = 2, 3, \ldots$. The last section contains our conclusions. The appendices are devoted to a discussion of technical details: in Appendix~\ref{app:C}, we derive the answer for two-dimensional conformal ladder integrals using the representation with Gegenbauer polynomials, in Appendix~\ref{app:d}, we give an explicit check of the loop shift identity for four-dimensional diagram with $\beta = 1$, Appendix~\ref{app:a} shows the properties of the answer in the arbitrary dimensional case with $\beta = 1$, and in Appendix~\ref{app:B}, we discuss the details of two-dimensional factorization.

\section{Conformal four-point
ladder integrals in diverse dimensions \label{sect:2}}

\subsection{Formulation of the problem and introductory remarks\label{sect:2.1}}

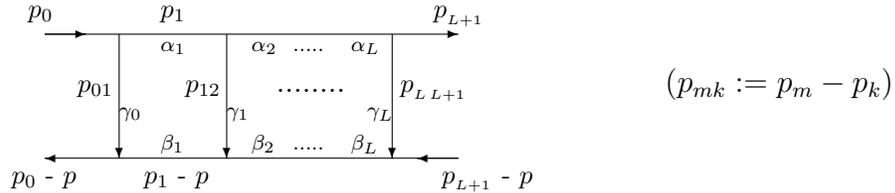
\begin{figure}[h]
\unitlength=11mm
\begin{picture}(25,2.5)(-3,0.5)

\put(1,2.5){\vector(1,0){5}}
\put(6,1){\vector(-1,0){5}}

\put(1,2.5){\vector(1,0){0.5}}
\put(6,1){\vector(-1,0){0.5}}

\put(1.9,2.5){\vector(0,-1){1.5}}
\put(3.2,2.5){\vector(0,-1){1.5}}
\put(5.2,2.5){\vector(0,-1){1.5}}
\put(3.8,1.8){$........$}

\put(0.8,2.7){\footnotesize $p_0$}
\put(0.6,0.7){\footnotesize $p_0$ - $p$}
\put(2.4,2.7){\footnotesize $p_1$}
\put(2.2,0.7){\footnotesize $p_1$ - $p$}
\put(5.7,2.7){\footnotesize $p_{_{L+1}}$}
\put(5.8,0.7){\footnotesize $p_{_{L+1}}$ - $p$}

\put(1.4,1.8){\footnotesize $p_{01}$}
\put(2.7,1.8){\footnotesize $p_{12}$}
\put(5.3,1.8){\footnotesize $p_{_{L\,L+1}}$}

\put(2.4,2.3){\scriptsize $\alpha_1$}
\put(3.5,2.3){\scriptsize $\alpha_2$}
\put(4,2.3){\scriptsize $.....$}
\put(4.7,2.3){\scriptsize $\alpha_L$}
\put(2.4,1.1){\scriptsize $\beta_1$}
\put(3.5,1.1){\scriptsize $\beta_2$}
\put(4,1.1){\scriptsize $.....$}
\put(4.7,1.1){\scriptsize $\beta_L$}
\put(1.9,1.5){\scriptsize $\gamma_0$}
\put(3.2,1.5){\scriptsize $\gamma_1$}
\put(4.9,1.5){\scriptsize $\gamma_L$}

\put(8.5,1.8){$(p_{mk} := p_m - p_k)$}


\end{picture}
\caption{\small\it The $L$-loop ladder diagram in momentum
space for massless $\phi^3$ theory.\label{fig1}}
\end{figure}

\begin{figure}[h]
\unitlength=7.5mm
\begin{picture}(24,8)(-3.5,-1)

\put(1,3){\line(1,0){12}}

\put(2.9,2.85){$\bullet$}
\put(4.9,2.85){$\bullet$}
\put(3.7,3.2){\footnotesize $\gamma_{_1}$}
\put(1.8,3.2){\footnotesize $\gamma_{_0}$}
\put(4,4.7){\footnotesize $\alpha_1$}
\put(4.5,0.9){\footnotesize $\beta_1$}

\put(5.6,3.2){\footnotesize $\gamma_{_2}$}
\put(6,4.6){\footnotesize $\alpha_2$}
\put(6.1,0.9){\footnotesize $\beta_2$}
\put(6.7,4.6){$\cdots$}
\put(6.5,3.2){$\cdots$}
\put(6.7,0.9){$\cdots$}

\put(5,3){\line(1,2){2}}
\put(3,3){\line(1,1){4}}
\put(9,3){\line(-1,2){2}}
\put(11,3){\line(-1,1){4}}

\put(5,3){\line(1,-2){2}}
\put(3,3){\line(1,-1){4}}
\put(9,3){\line(-1,-2){2}}
\put(11,3){\line(-1,-1){4}}

\put(9.1,0.8){\scriptsize $\beta_L$}
\put(9.3,4.7){\scriptsize $\alpha_L$}
\put(9.6,3.4){\scriptsize $\gamma_{_{L-1}}$}
\put(11.5,3.4){\scriptsize $\gamma_{_L}$}

\put(8.8,2.85){$\bullet$}
\put(10.8,2.85){$\bullet$}

\put(6.8,7.2){\footnotesize $x_2$}
\put(0.5,2.5){\footnotesize $x_1$}
\put(12.7,2.5){\footnotesize $x_3$}
\put(6.5,-1.3){\footnotesize  $x_4$}

\put(2.8,2.6){\footnotesize $x_{_5}$}
\put(4.7,2.6){\footnotesize $x_{_6}$}
\put(7.9,2.7){\footnotesize $x_{_{L+3}}$}
\put(9.8,2.8){\scriptsize $x_{_L+4}$}

\put(2.5,2){\dashbox{0.1}(9,2)[sb]{}}
 \multiput(9.6,2)(0,0.24){9}{\line(0,1){0.1}}

  \multiput(4.3,2)(0,0.24){9}{\line(0,1){0.1}}
   \multiput(6.2,2)(0,0.24){9}{\line(0,1){0.1}}

     \multiput(1,2)(0.24,0){9}{\line(1,0){0.11}}
   \multiput(1,4)(0.24,0){7}{\line(1,0){0.11}}

   \multiput(11.5,2)(0.24,0){7}{\line(1,0){0.11}}
   \multiput(11.5,4)(0.24,0){7}{\line(1,0){0.11}}

\end{picture}
\caption{\label{fig2}\small\it Dual graphical representation of the integral
(\ref{br1}) written in the form (\ref{genf07}).
 The integrations are performed over the boldface vertices
and each line with index $\alpha$:
 $x_i \stackrel{\alpha}{-\!\!\!-\!\!\!-\!\!\!-}  x_j$,
corresponds to the propagator
$x_{ij}^{-2\alpha}$}
\end{figure}
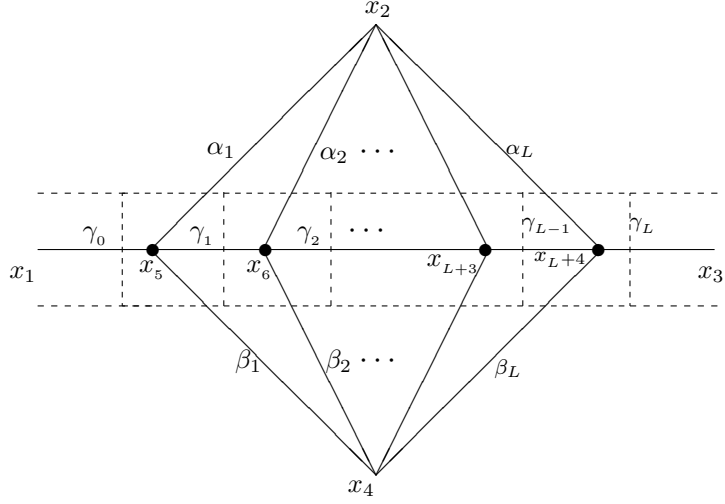

To fix notation, in this section we recall some known facts (see e.g. Refs. \cite{Isa, DIS2}).
 For massless $\phi^3$ theory, we consider
 D-dimensional $L$-loop ladder  integrals
 with arbitrary indices on the lines
$\alpha_k,\beta_k,\gamma_k$:
\be
\lb{br1}
D_L (p_0, p_{L+1}, p) =
\int \left[ \prod_{k=1}^{L}
\frac{d^D p_k}{p^{2\alpha_k}_k \, (p_k - p)^{2\beta_k}} \right]
\prod_{m=0}^{L} \frac{1}{(p_{m+1} - p_m)^{2\gamma_m}} \; .
\ee
 These integrals
correspond to the momentum-space Feynman diagrams
depicted in Fig.~\ref{fig1} (here the integrations are performed
over the loop momenta $p_i$ ($i=1,...,L$)). The diagram given in Fig.~\ref{fig1}
is presented in the dual form in Fig.~\ref{fig2}.
Here the integrations are performed over the boldface vertices
which are placed in the boxes of the diagram  in Fig.~\ref{fig1}.

\noindent
The $L$-loop ladder integral which corresponds to the dual
diagram in Fig.\ref{fig2} is written as
\be
\lb{genf07}
\begin{array}{c}
I^{(L)}(x_1,x_2,x_3,x_4;\alpha_i,\beta_i\gamma_i) =
\int d^Dx_5 \dots d^D x_{L+4} \; \times \\ [0.3cm]
\displaystyle
\times \; \left. \frac{1}{(x_{1,5})^{2\gamma_0}}\,
\prod\limits_{i=1}^{L} \, \frac{1}{(x_{2,i+4})^{2\alpha_i}}\,
 \frac{1}{(x_{4,i+4})^{2\beta_i}}\,
 \frac{1}{(x_{i+4,i+5})^{2\gamma_i}}
 \right|_{x_{_{L+5}} \equiv x_3} \; ,
\end{array}
\ee
and $I^{(0)}=x_{13}^{-2\gamma_0}$.
To identify the integrals (\ref{br1}) and (\ref{genf07}),
we relate the variables
$$
\begin{array}{c}
x_{12} = p_0 \; , \;\;\; x_{23} = -p_{_{L+1}} \; , \;\;\;
x_{34} = p_{_{L+1}} - p \; , \;\;\;
x_{41} = p - p_0  \; , \\ [0.2cm]
p_{i,i+1}= x_{i+4,i+5} \; , \;\;\; p_i = x_{i+4,2}
\; , \;\;\; p_i -p = x_{i+4,4}  \;\;\;\; (i=1,2,...) \; ,
\end{array}
$$
and one can
choose $x_2=0$ using translation invariance of the integral
(\ref{genf07}).

For further applications, we fix the indices in the integral (\ref{genf07}),
which is depicted in Fig.\ref{fig2},
by the conformal conditions for all boldface vertices
\cite{Isa, DIS2}
\be
\lb{genf08}
\gamma_{k-1} + \alpha_k +\beta_k + \gamma_k = D \; .
\ee
When $\alpha_k,\beta_k,\gamma_k$ are positive real
numbers, the condition (\ref{genf08})
ensures the convergence of the integral (\ref{genf07}).
Under the conformal transformations
$y_\mu \to \frac{y_\mu}{y^2}$ (inversion of the vector
$y \in \mathbb{R}^{D}$; $\mu=1,...,D$ is the space index)  we have
for vectors $x_i,x_j \in \mathbb{R}^{D}$:
$$
d^Dx_i  \; \to \; \frac{d^Dx_i }{x_i^{2D}} \; , \;\;\;\;\;
(x_{ij})^2 \; \to \; \frac{(x_{ij})^2}{x_i^{2} x_j^{2}} \; ,
$$
and in the case when the conditions (\ref{genf08}) are
fulfilled, the integral
(\ref{genf07}) transforms as follows:
\be
\lb{genf09}
\begin{array}{c}
 I^{(L)}(\frac{1}{x_i};\alpha_i,\beta_i,\gamma_i) =
 (x_1)^{2\gamma_0} (x_2)^{2A} (x_3)^{2\gamma_{L}}
 (x_4)^{2B}
 I^{(L)}(x_i;\alpha_i,\beta_i,\gamma_i) \\ [0.2cm]
 \boxed{ \dfrac{1}{x_i} :=  \dfrac{x_{i \mu}}{(x_i)^2} } \; , \;\;\;\;\;
 A=\sum_i \alpha_i \; , \;\;\;\;\; B=\sum_i \beta_i \; .
 \end{array}
\ee
Thus,
the function\footnote{We use the notation $(x_{ij})^A= (x_{ij}^2)^{A/2}$.}
$$
(x_{24})^{2 A} (x_{13})^{(\gamma_0+A-B+\gamma_L)}
(x_{34})^{(-\gamma_0-A+B + \gamma_L)}
(x_{14})^{(\gamma_0-A+B-\gamma_L)}
 I^{(L)}(x_i;\alpha_i,\beta_i,\gamma_i) \; ,
$$
is invariant under all conformal transformations
(the invariance of this function under
Poincar\'{e} and scale 
transformations is obvious)
and therefore is expressed as a function of two cross-ratios
 \be
 \lb{crossr}
u = \frac{x_{12}^2 x_{34}^2}{x_{24}^2 x_{13}^2}\; , \;\;\;\;
 v  = \frac{x_{14}^2 x_{23}^2}{x_{24}^2 x_{13}^2} \; ,
 \ee
so we have
\be
\lb{genf0}
 x_{24}^{2 A} x_{13}^{(\gamma_0+A-B+\gamma_L)}
x_{34}^{(-\gamma_0-A+B + \gamma_L)}
x_{14}^{(\gamma_0-A+B-\gamma_L)}
 I^{(L)}(x_i;\alpha_i,\beta_i,\gamma_i) =
 f_{(L)}(u,v;\alpha_i,\beta_i,\gamma_i) \; ,
\ee
and $f_{(0)} = 1$. Further, for simplicity we fix the parameters
$\beta_k=D/2-\gamma_k
\equiv \beta$ (which gives $\alpha_k=\beta$).
In this case,
formula (\ref{genf0}) is simplified to
\be
\lb{genf00}
(x_{24})^{2 L \beta} (x_{13})^{2(D/2-\beta)}
 I^{(L)}(x_i;\beta) \equiv f_{(L)}(u,v;\beta) \; ,
\ee
and these functions (associated with the diagrams in Fig~\ref{fig2}) give contributions to the $4$-point amplitude in general $D$-dimensional bi-scalar fishnet CFT proposed in~\cite{KO, Kazakov:2022dbd}. For this choice of the
parameters,  we found in \cite{Isa, Isaev:2007uy, DIS2} that
 the generating function of the $L$-loop ladder integrals
 (\ref{genf07}) (with special normalization
 (\ref{genf00})) is represented in the form
 of a Green's function for conformal
 quantum mechanics\footnote{Here to avoid confusion with the standard notation of cross-ratios, we use the notation ${\sf x}$ and ${\sf y}$ instead of $u$ and $w$ in \cite{DIS2} and $u$ and $v$ in \cite{Isa}.}
 \be
\lb{genf01i}
\begin{array}{c}
 \displaystyle
\frac{a(\beta)}{({\sf x} - {\sf y})^{2(D/2-\beta)}}  \cdot
\sum_{L=0}^\infty
\bigl(g \, a(\beta) \bigr)^{L} \, f_{(L)}(u,v;\beta) =
\bigl\langle {\sf x}\bigr| \, \frac{1}{\hat{p}^{2\beta} -
g \, \hat{q}^{-2\beta}} \,
\bigl| {\sf y}\bigr\rangle \; , \\ [0.2cm]
\hat{p}^{2} = \hat{p}_\mu \hat{p}_\mu \, , \;\;\;
\hat{q}^{2} = \hat{q}_\mu \hat{q}_\mu \, , \;\;\;
[\hat{q}_\mu , \; \hat{p}_\nu] = i \delta_{\mu\nu} \, ,
\;\;\;\; {\sf x} := \frac{1}{x_{12}} - \frac{1}{x_{42}} \, ,
\;\;\;\; {\sf y} := \frac{1}{x_{32}} - \frac{1}{x_{42}}  \, , \\ [0.3cm]
\dfrac{a(\beta)}{(x - y)^{2(D/2-\beta)}} =
\langle x| \dfrac{1}{\hat{p}^{2 \beta} } |y \rangle \,\quad (\forall x,y \in \mathbb{R}^D) , \;\;\; \;\;\;
  a(\beta):=
 \dfrac{\Gamma(D/2-\beta) }{ 2^{2 \beta} \, \pi^{D/2} \Gamma(\beta) } \; ,
 \end{array}
\ee
 where $\{\hat{q}_\mu , \; \hat{p}_\nu\}$ are the generators
 of $D$-dimensional Heisenberg algebra,
 and \cite{DIS2}
\be
\lb{genf01}
\begin{array}{c}
\bigl\langle {\sf x}\bigr| \, \frac{1}{\hat{p}^{2\beta} -
g \, \hat{q}^{-2\beta}} \,
\bigl| {\sf y}\bigr\rangle
\displaystyle =
{\sf x}^{2\beta} \,
\sum\limits_{n=0}^\infty \mu(n) \, \frac{
{\sf x}^{\mu_1 ... \mu_n}\, {\sf y}^{\mu_1 ... \mu_n}}{
({\sf x}^2\, {\sf y}^2)^{(D/4+n/2)}}
\int\limits_{-\infty}^{+\infty} d\nu
 \, \frac{({\sf y}^2/{\sf x}^2)^{{\sf i}\nu}}{(\tau_{n,\nu}(\beta)- g)} = \\ [0.3cm]
 \displaystyle =
 {\sf x}^{2\beta} \, \sum\limits_{L=0}^\infty g^L
\sum\limits_{n=0}^\infty \mu(n) \, \frac{
{\sf x}^{\mu_1 ... \mu_n}\, {\sf y}^{\mu_1 ... \mu_n}}{
({\sf x}^2\, {\sf y}^2)^{(D/4+n/2)}}
\int\limits_{-\infty}^{+\infty} d\nu
 \, \frac{({\sf y}^2/{\sf x}^2)^{{\sf i}\nu}}{
 (\tau_{n,\nu}(\beta))^{L+1}} \;
\end{array}
\ee
Here ${\sf x}^{\mu_1...\mu_n}$ is the traceless symmetric
tensor
with components that are homogeneous
in ${\sf x}^\mu$ polynomials
 (see Appendix A in \cite{DIS2}); the functions
 \be
\lb{wfunc}
\mu(n) = \frac{2^{n-1} \Gamma(D/2+n)}{\pi^{D/2+1} n!} \;
 \ee
  were introduced in \cite{DIS2} as weights
in the completeness condition
of the eigenfunctions of operators
$H_\beta = \hat{p}^{2\beta}\hat{q}^{2\beta}$;
the operators $H_\beta$
form a commutative set
for all $\beta$ \cite{Isa} while $\tau_{n,\nu}(\beta;\lambda)$ are eigenvalues of $H_\beta$ (for further details see subsection~\ref{sect:2.3}):
\be
\lb{tau1}
\tau_{n,\nu}(\beta;\lambda) =
4^\beta \frac{\Gamma(\frac{\lambda+n+1}{2}+\beta -{\sf i}\nu)}{
\Gamma(\frac{\lambda+n+1}{2}-\beta + {\sf i}\nu)} \;
\frac{\Gamma(\frac{\lambda+n+1}{2}+{\sf i}\nu)}{
\Gamma(\frac{\lambda+n+1}{2}-{\sf i}\nu)}
 \; , \;\;\;\;\;
\boxed{ \lambda := \frac{D-2}{2}} \; .
\ee
Note that
\be
\lb{genf03}
\tau_{n,\nu}(\beta) =  4{\textstyle \left(\frac{\lambda+n+1}{2}+{\sf i}\, \nu - \beta\right)
\left(\frac{\lambda+n-1}{2}+\beta - {\sf i}\, \nu\right)}\,
\tau_{n,\nu}(\beta-1)  \; .
\ee
and for integer $\beta = k \in \mathbb{Z}_{>0}$ we obtain
 \be\label{genf03a}
 \tau_{n,\nu}(k) =  4^k \prod_{m=1}^{k}
 \textstyle\left(\frac{\lambda+n+1}{2}+{\sf i}\, \nu - m\right)
 \left(\frac{\lambda+n-1}{2}+m - {\sf i}\,\nu\right) \,.
\ee
Note also that we have relations between the variables ${\sf x},{\sf y}$
(introduced in (\ref{genf01i})) and
cross-ratios (\ref{crossr})
\be
\lb{genf03b}
\frac{{\sf y}^2}{{\sf x}^2} = \frac{u}{v} \; , \;\;\;\;\;
\frac{({\sf x}-{\sf y})^2}{{\sf x}^2} = \frac{1}{v} \; .
\ee

Now we consider the expansion
 \be
\lb{genf04}
\bigl\langle {\sf x}\bigr| \,
\frac{1}{\hat{p}^{2\beta} -
g \, \hat{q}^{-2\beta}} \,
\bigl| {\sf y}\bigr\rangle =
 \sum^\infty_{L=0}
\frac{1}{L!} \left({g \over 4} \right)^L \,
\Phi^{(\beta)}_L({\sf x},{\sf y}) \; ,
\ee
where $\Phi^{(\beta)}_L$ is related (in view of (\ref{genf01i}))
 to the integral (\ref{genf07}) with
 $\alpha_i=\beta_i=D/2-\gamma_i\equiv\beta$
 (in Fig.~\ref{fig1} the horizontal and vertical propagators
 have indices $\beta$ and $D/2-\beta$, respectively):
 \be
\lb{genf04b}
 \Phi^{(\beta)}_L({\sf x},{\sf y}) =\frac{L!4^L a(\beta)^{L+1}}{
 ({\sf x}-{\sf y})^{2(D/2-\beta)}} f_{(L)}=
 L!4^L a(\beta)^{L+1} x_{12}^{2(D/2-\beta)} x_{23}^{2(D/2-\beta)}
 x_{24}^{2L\beta} I^{(L)} \; .
 \ee
The representation (\ref{genf01i}) of the generating function of
$L$-loop ladder integrals
as a Green's function for conformal quantum mechanics
is very useful since we, for example, immediately
deduce
 the symmetry properties of
 $\Phi^{(\beta)}_L({\sf x},{\sf y})$ \cite{Isa}:
 \be
 \lb{genf04d}
\Phi_L^{(\beta)}({\sf x},{\sf y}) = \Phi_L^{(\beta)}({\sf y},{\sf x}) =
({\sf x}^2 {\sf y}^2)^{(\beta - {D \over 2})} \,
\Phi_L^{(\beta)} \Bigl({1 \over {\sf x}}, {1 \over {\sf y}}\Bigr)  \; ,
 \ee
which also follows from (\ref{genf01}).
Comparing (\ref{genf01}) and (\ref{genf04}) we obtain \cite{DIS2}
 \be
\lb{genf05}
\Phi^{(\beta)}_L({\sf x},{\sf y})  =
 {\sf x}^{2\beta} \, (L! \, 4^L)
\sum\limits_{n=0}^\infty \mu(n) \, \frac{
{\sf x}^{\mu_1 ... \mu_n}\, {\sf y}^{\mu_1 ... \mu_n}}{
({\sf x}^2\, {\sf y}^2)^{(D/4+n/2)}}
\int\limits_{-\infty}^{+\infty} d\nu
 \, \frac{({\sf y}^2/{\sf x}^2)^{{\sf i}\nu}}{(\tau_{n,\nu}(\beta;\lambda))^{L+1}} \, .
\ee

\vspace{0.3cm}
\noindent
{\bf Remark 1}. 
Under the translation of the variable
 $\nu \mapsto \nu - {\sf i}\frac{\beta}{2}$,
the function~\eqref{genf05}
transforms to the expression
\begin{align}
\label{nu-sim}
    \Phi^{(\beta)}_L({\sf x},{\sf y})  =& \, (L! \, 4^L) 4^{-\beta(L + 1)}
    \sum\limits_{n=0}^\infty \mu(n) \, \frac{
    {\sf x}^{\mu_1 ... \mu_n}\, {\sf y}^{\mu_1 ... \mu_n}}{
    ({\sf x}^2\, {\sf y}^2)^{(D/4+n/2)}} \quad\times \notag \\ &\quad\times
   \int\limits_{-\infty}^{+\infty} d\nu
    \, \dfrac{\Gamma^{L + 1}(\frac{\lambda+n+1}{2} - \frac{\beta}{2} - {\sf i}\nu)\Gamma^{L + 1}(\frac{\lambda+n+1}{2} - \frac{\beta}{2} + {\sf i}\nu)}{\Gamma^{L + 1}(\frac{\lambda+n+1}{2} + \frac{\beta}{2} - {\sf i}\nu)\Gamma^{L + 1}(\frac{\lambda+n+1}{2} + \frac{\beta}{2} + {\sf i}\nu)} ({\sf y}^2)^{{\sf i}\nu + \beta/2}({\sf x}^2)^{-{\sf i}\nu + \beta/2}\, ,
\end{align}
which is evidently real and symmetric with respect to ${\sf x} \leftrightarrow {\sf y}$. Note that under the translation $\nu \mapsto \nu - \iu\frac{\beta}{2}$ the contour of integration shifts to $\Im \nu = \frac{\beta}{2}$. It is possible to move it back to $\Re\nu = 0$ and obtain~\eqref{nu-sim} if $\beta < \lambda + 1$, so no poles lie in the band $0 < \Im \nu < \frac{\beta}{2}$.
\vspace{0.2cm}

\noindent
\textbf{Remark 2}. For the case of generic $\beta$ and $D$ we denote the contour of integration in~\eqref{genf05} as $\Im \nu =0$. However, one should note that in the special cases one should understand it as a contour, which separates the two series of singularities arising from the gamma functions in the definition of $\tau_{n,\nu}(\beta; \lambda)$ (namely, one series at $\nu = \iu(\frac{\lambda + n + 1}{2} - \beta) + \iu k$ and the other at $\nu = -\iu\frac{\lambda + n + 1}{2} - \iu k$, where $k \in \mathbb{Z}_{\ge 0}$). For more details in the cases of specific $\beta$ and $D$, see section~\eqref{sect:3.2} and Appendix~\ref{app:B}.

\vspace{0.2cm}
Now we use
the definition of the Gegenbauer polynomials
$C_n^{(D/2-1)}$ \cite{Kot1, Kot2}
(see also \cite{DIS2}, eq. (3.48)) in (\ref{genf05}):
 \be
 \lb{derk01}
{\sf x}^{\mu_1...\mu_n} {\sf y}^{\mu_1...\mu_n} = \frac{n! \Gamma(D/2-1)}{2^n \Gamma(n+D/2-1)}
C_n^{(D/2-1)}(\hat{{\sf x}}\hat{{\sf y}}) ({\sf x}^2 {\sf y}^2)^{n/2} \; ,
 \ee
where $\hat{{\sf x}}\hat{{\sf y}} =
\frac{({\sf x}{\sf y})}{\sqrt{{\sf x}^2 {\sf y}^2}}$.
One can consider the right-hand side of this formula as a correct
analytical continuation of its left-hand side for any non-integer $D$.
Let us introduce the notation $\lambda = \frac{D}{2}-1$
and new parametrization \cite{Isa}
\begin{align}
\label{genf11}
\frac{{\sf y}^2}{{\sf x}^2} = z\bar{z} \ \ ;\ \
\frac{2({\sf x}{\sf y})}{{\sf x}^2} = z+ \bar{z} \quad \Rightarrow \quad
\hat{{\sf x}}\hat{{\sf y}} = \frac{z+\bar{z}}{2\sqrt{z\bar{z}}}
 \; .
\end{align}
Note that in terms of conformal ratios~\eqref{crossr}
and in view of (\ref{genf03b}), this parametrization corresponds to
\begin{align}
    u = \dfrac{z\bar{z}}{(1 - z)(1 - \bar{z})} \; ,
     \quad \quad  v = \dfrac{1}{(1 - z)(1 - \bar{z})} \; .
\end{align}

Using definition (\ref{derk01}), formula
(\ref{wfunc}) and parametrization (\ref{genf11}),
we write (\ref{genf05}) as
\begin{align}\label{beta}
\Phi^{(\beta)}_L({\sf x},{\sf y})
=  \dfrac{\Gamma(\lambda) \, L! \, 4^L  {\sf x}^{2\beta}}{
2 \pi^{\lambda+2}({\sf x}^2\, {\sf y}^2)^{D/4}} \,
\sum\limits_{n=0}^\infty
(n+\lambda)\,
C^{(\lambda)}_n\left(\dfrac{z+\bar{z}}{2\sqrt{z\bar{z}}}\right)\,
\int\limits_{-\infty}^{+\infty} d\nu
 \dfrac{(z\bar{z})^{{\sf i}\nu}}{\left[\tau_{n,\nu}(\beta;\lambda)\right]^{L+1}}
\end{align}
 The case $D=2$ requires special consideration and
there are two possible approaches. One is to perform the limit $D \to 2$ ($\lambda \to 0$) in~\eqref{beta} using the relation between the Gegenbauer and Chebyshev polynomials. We discuss this method in Appendix~\ref{app:C}. Another thing we adopt here is the introduction of the complex coordinates
 $x = {\sf x}_1+\iu{\sf x}_2\,, \bar{x} = {\sf x}_1-\iu{\sf x}_2$ and
$y = {\sf y}_1+\iu{\sf y}_2\,, \bar{y} = {\sf y}_1-\iu{\sf y}_2$, in~\eqref{genf05}.
Indeed, for $D=2$ and $n \geq 1$ we have
\begin{align}
\label{D2}
{\sf x}^{\mu_1...\mu_n} {\sf y}^{\mu_1...\mu_n} =
\frac{1}{2^n}\left(x^n\bar{y}^n+\bar{x}^ny^n\right)
\ \ ;\ \ {\sf x}^2 = x\bar{x}\ \ ;\ \ {\sf y}^2 = y\bar{y}
\ \ ;\ \ ({\sf x}{\sf y}) = \frac{1}{2}\left(x\bar{y}+\bar{x}y\right).
\end{align}
The parametrization \eqref{genf11} is reduced to the form
\begin{align}
\frac{y\bar{y}}{x\bar{x}} = z\bar{z} \ \ ;\ \
\frac{y}{x}+\frac{\bar{y}}{\bar{x}} = z + \bar{z}
\quad \Rightarrow z = \frac{y}{x} \ \ ;\ \ \bar{z} = \frac{\bar{y}}{\bar{x}}
\end{align}
so that due to \eqref{D2} one obtains
\begin{align}
\frac{
    {\sf x}^{\mu_1 ... \mu_n}\, {\sf y}^{\mu_1 ... \mu_n}}{
    ({\sf x}^2\, {\sf y}^2)^{n/2}} = \frac{1}{2^n}\frac{z^n+\bar{z}^n}{(z\bar{z})^{n/2}} \quad \text{for }n \ge 1, &&  \frac{
    {\sf x}^{\mu_1 ... \mu_n}\, {\sf y}^{\mu_1 ... \mu_n}}{
    ({\sf x}^2\, {\sf y}^2)^{n/2}} \overset{n = 0}{=} 1.
\end{align}
In $D=2$ expression \eqref{wfunc} for $\mu(n)$ looks simpler
\begin{align*}
\left.\mu(n)\right|_{D=2} = \frac{2^{n-1}}{\pi^{2}} \;
\end{align*}
and finally the general formula \eqref{genf05} is reduced in $D=2$
to the following form:
\begin{multline}
\label{D=2}
\left. \Phi^{(\beta)}_L({\sf x},{\sf y}) \right|_{\lambda=0} \; = \;
\dfrac{ L! \, 4^L  {\sf x}^{2\beta}}{
2 \pi^{2}({\sf x}^2\, {\sf y}^2)^{1/2}} \\
\left[\int\limits_{-\infty}^{+\infty} d\nu
\, \frac{(z\bar{z})^{{\sf i}\nu}}{(\tau_{0,\nu}(\beta;0))^{L+1}} + \sum\limits_{n=1}^\infty \left(z^n+\bar{z}^n\right)
\int\limits_{-\infty}^{+\infty} d\nu
 \, \frac{(z\bar{z})^{{\sf i}\nu-n/2}}{(\tau_{n,\nu}(\beta;0))^{L+1}}\right] \, .
\end{multline}
In Appendix \ref{app:C} we show that by means
of the symmetry property $\tau_{n,\nu}(\beta;0) =
\tau_{-n,\nu}(\beta;0)$ one can write (\ref{D=2})
in the concise form (see (\ref{D=2b}))
\begin{equation}
\lb{D=2d}
\left. \Phi^{(\beta)}_L({\sf x},{\sf y}) \right|_{\lambda=0} \; = \;
\dfrac{ L! \, 4^L  {\sf x}^{2\beta}}{
2 \pi^{2}({\sf x}^2\, {\sf y}^2)^{1/2}} \left[
 \sum_{n \in \mathbb{Z}}\;
\int\limits_{-\infty}^{+\infty} d\nu
\dfrac{z^{{\sf i}\nu+\frac{n}{2}}\,\bar{z}^{{\sf i}\nu-\frac{n}{2}}}
 {\left[\tau_{n,\nu}(\beta;0)\right]^{L+1}} \right] \,.
\end{equation}

\vspace{0.5cm}

\noindent
{
{\bf Remark 3.} The symmetries (\ref{genf04d}), in view of (\ref{genf03b}) and (\ref{genf04b}), are equivalent to
$f_{(L)}(u,v;\beta) = f_{(L)}(v,u;\beta)$
and in terms of
$z$ and $\bar{z}$ are written as
\be
\lb{symt}
f_{(L)}(z,\bar{z};\beta) =
f_{(L)}(1/z,1/\bar{z};\beta) \; ,
\ee
(these symmetries for $L=1$, $\beta=1$ were exploited
 in \cite{Loebb2}).}

\vspace{0.5cm}

\noindent
{\bf Remark 4.}
It is well known~\cite{Symanzik:1972wj} that the $n$-point conformal vertex, for which
the sum of indices on the lines satisfies
\be
\lb{genf15}
\sum_{i=1}^n \beta_i =D \; ,
\ee
transforms under conformal inversions
$x_i \to \frac{1}{x_i}$ as follows:

\unitlength=3.1mm
\begin{picture}(25,11)(-3,0)

\put(8,4){\line(-1,1){3.5}}
\put(8,4){\line(1,1){3.5}}
\put(8,4){\line(-3,-2){4}}
\put(8,4){\line(3,-2){4}}

\put(7.75,3.75){$\bullet$}
\put(7,1.2){\large $\dots$}

\put(5,3){\scriptsize $\beta_n$}
\put(6,6.7){\scriptsize $\beta_1$}
\put(10,5){\scriptsize $\beta_2$}
\put(9.5,1.5){\scriptsize  $\beta_3$}

\put(2.2,0.8){\small $\frac{1}{x_n}$}
\put(4,8.7){\small $\frac{1}{x_1}$}
\put(12,7){\small $\frac{1}{x_2}$}
\put(11.8,0){\small  $\frac{1}{x_3}$}

\put(15,3.8){$ \;\; = \;\;\;\; \prod\limits_{i=1}^n x_i^{2\beta_i}$}


\put(28,4){\line(-1,1){3.5}}
\put(28,4){\line(1,1){3.5}}
\put(28,4){\line(-3,-2){4}}
\put(28,4){\line(3,-2){4}}

\put(27.75,3.75){$\bullet$}
\put(27,1.2){\large $\dots$}

\put(25,3){\scriptsize $\beta_n$}
\put(26,6.7){\scriptsize $\beta_1$}
\put(30,5){\scriptsize $\beta_2$}
\put(29.5,1.5){\scriptsize  $\beta_3$}

\put(22,1){\small $x_n$}
\put(24,8.7){\small $x_1$}
\put(32,7){\small $x_2$}
\put(31.5,0){\small  $x_3$}

\end{picture}

\vspace{0.5cm}

\noindent
It is evident that the
 integral $I(x_1,...,x_n;\vec{\beta})$,
which is depicted as a
Feynman graph with $n$ external lines having indices $\vec{\beta}=(\beta_1,...,\beta_n)$
and with all internal boldface vertices being conformal,
transforms under inversions exactly as a conformal $n$-point vertex:
 $I(\frac{1}{x_1},...,\frac{1}{x_n};\vec{\beta})) =
 \prod\limits_{i=1}^n x_i^{2\beta_i} I(x_1,...,x_n;\vec{\beta}))$.
 Then, a conformal invariant function can be chosen as
 $$
 f(u^{ij}_{kl},...) = I(x_1,...,x_n;\vec{\beta}) \prod_{i=1}^n
 \dfrac{(x_{i,i+1})^{\beta_i} (x_{i,i+2})^{\beta_i}}{(x_{i+1,i+2})^{\beta_i}}
 \; , \;\;\;\;\;\; (n+i=i) \; ,
 $$
 where the conformal
 cross-ratios are $u^{ij}_{kl} = x_{ij}^2 x_{kl}^2/(x_{ik}^2 x_{jl}^2)$
 and $(i,j,k,l)$ are all possible
 different 4-element subsets of the $n$ element set
 $(1,2,...,n)$.
 Formulas (\ref{genf09}), (\ref{genf0}) are examples of these
 general rules.

\vspace{0.5cm}

\noindent
{\bf Remark 5.} Note that one
can shift the indices on the lines
by one unit in the integrals (\ref{br1}),
(\ref{genf07}) by making use of the generalized integration-by-parts formula
for the $n$-point vertex (see \cite{GorIs})

\unitlength=3.5mm
\begin{picture}(25,16)(1,-8)

\put(0,3){\line(1,0){6}}
\put(3,0){\line(0,1){6}}

\put(2.75,2.75){$\bullet$}
\put(1.3,1.5){\large $\ddots$}

\put(1,3.4){\scriptsize $\beta_n$}
\put(3.3,4.5){\scriptsize $\beta_1$}
\put(4.5,3.3){\scriptsize $\beta_2$}
\put(3.3,0.8){\scriptsize  $\beta_3$}

\put(6.8,2.8){$= \; \dfrac{1}{D-\beta_1 - \sum\limits_{i=1}^n \beta_i}
{\Large \Bigl(} \! {\scriptsize  \beta_2} \large \Bigl( $}


\put(20,3){\line(1,0){4}}
\put(22,1){\line(0,1){4}}

\put(21.75,2.75){$\bullet$}
\put(20.5,1.7){\large $\ddots$}

\put(19,2.5){\scriptsize $\beta_n$}
\put(21.7,5.6){\scriptsize $\beta_1$-$1$}
\put(24.5,2.5){\scriptsize $\beta_2$}
\put(21,0.3){\scriptsize  $\beta_3$}

\put(25.8,2.8){$-$}


\put(28,3){\line(1,0){4}}
\put(30,1){\line(0,1){4}}
\put(30,5){\line(1,-1){2}}

\put(29.75,2.75){$\bullet$}
\put(28.5,1.7){\large $\ddots$}

\put(27,2.5){\scriptsize $\beta_n$}
\put(29.7,5.5){\scriptsize $\beta_1$}
\put(32.2,2.5){\scriptsize $\beta_2$}
\put(30,0){\scriptsize  $\beta_3$}
\put(31.2,4){\scriptsize  -$1$}

\put(33,2.8){${\large \Bigr)} + \beta_3
{\large \Bigl( \; . \; . \; . \; \Bigr)} + \dots $}


\put(20,-4){$\dots + {\scriptsize  \beta_n} \large \Bigl( $}

\put(26,-4){\line(1,0){4}}
\put(28,-6){\line(0,1){4}}

\put(27.75,-4.25){$\bullet$}
\put(26.5,-5.3){\large $\ddots$}

\put(25,-4.5){\scriptsize $\beta_n$}
\put(27.7,-1.5){\scriptsize $\beta_1$-$1$}
\put(30.5,-4.5){\scriptsize $\beta_2$}
\put(28,-7){\scriptsize  $\beta_3$}

\put(31.8,-4.2){$-$}


\put(34,-4){\line(1,0){4}}
\put(36,-6){\line(0,1){4}}
\put(36,-2){\line(-1,-1){2}}

\put(35.75,-4.25){$\bullet$}
\put(34.5,-5.3){\large $\ddots$}

\put(33,-4.5){\scriptsize $\beta_n$}
\put(35.7,-1.5){\scriptsize $\beta_1$}
\put(38.2,-4.5){\scriptsize $\beta_2$}
\put(36,-7){\scriptsize  $\beta_3$}
\put(34,-3){\scriptsize  -$1$}

\put(39,-4.2){${\large \Bigr)} \Large \Bigr)$,}

\end{picture}

\noindent
where the small $\beta_i$ are the indices on the lines.
This equation is written for the case of a selected line
with index $\beta_1$. By selecting other lines with
indices $\beta_i$, one obtains other similar identities.
Note that if we use
these identities (in the case $n=4$) for the integral (\ref{genf07})
(presented in Fig. \ref{fig2}), we obtain a sum of
 integrals with shifted indices of lines, i.e.
 the conformal vertex conditions (\ref{genf15})
 are broken for certain boldface vertices.


\subsection{Dimensional shift for generic
propagator index $\beta$\label{sect:2.2}}

{The relations of (conformal)
Feynman integrals in different
dimensions (by means of a special operator acting
on the variables of the external legs)
were considered in many papers; see e.g.~\cite{DHP,Taras,Lee, VolSpr, Loebb,Petkou0, Petkou, Petkou2}.}

 Let us redefine (\ref{genf04b}), (\ref{beta})
 and introduce the function, which depends only on the
conformal kinematic parameters $z,\bar{z}$:
\begin{align}
\widetilde{\Phi}^{(\beta)}_L(z,\bar{z};\lambda)  &:=
\dfrac{2 \pi^{\lambda+2}({\sf x}^2\, {\sf y}^2)^{(\lambda+1)/2}}{
(z \bar{z})^{\lambda/2} \, {\sf x}^{2\beta} \, (L! \, 4^L)}\,
\Phi^{(\beta)}_L({\sf x},{\sf y}) \notag \\
 &= a(\beta)^{L+1} 2 \pi^{\lambda+2}
\dfrac{(z\bar{z})^{1/2}}{
((1-z)(1-\bar{z}))^{\lambda-\beta+1}}
\, f_L(z,\bar{z};\beta) \; ,
\label{beta0}
\end{align}
For this function, in view of
(\ref{beta}) and (\ref{D=2}), (\ref{D=2d}), we have representations
\begin{align}
\widetilde{\Phi}^{(\beta)}_L(z,\bar{z};\lambda)  & =
\dfrac{\Gamma(\lambda)}{(z \bar{z})^{\lambda/2}} \,
\sum\limits_{n=0}^\infty
(n+\lambda)\,
C^{(\lambda)}_n
\left(\dfrac{z+\bar{z}}{2\sqrt{z\bar{z}}}\right)\,
\int\limits_{-\infty}^{+\infty} d\nu
 \dfrac{(z\bar{z})^{{\sf i}\nu}}{\left[\tau_{n,\nu}(\beta;\lambda)\right]^{L+1}} \; ,
  \label{D=D} \\
\widetilde{\Phi}^{(\beta)}_L(z,\bar{z};0)  & =
 \int\limits_{-\infty}^{+\infty} d\nu
 \dfrac{(z\bar{z})^{{\sf i}\nu}}{
 \left[\tau_{0,\nu}(\beta;0)\right]^{L+1}}+
 \sum\limits_{n=1}^\infty \left(z^n+\bar{z}^n\right)\,
 \int\limits_{-\infty}^{+\infty} d\nu
 \dfrac{(z\bar{z})^{{\sf i}\nu-n/2}}{\left[\tau_{n,\nu}(\beta;0)
 \right]^{L+1}}   \label{D=2a} \\
 & =  \sum_{n \in \mathbb{Z}}\;
\int\limits_{-\infty}^{+\infty} d\nu
\dfrac{z^{{\sf i}\nu+\frac{n}{2}}\,\bar{z}^{{\sf i}\nu-\frac{n}{2}}}
 {\left[\tau_{n,\nu}(\beta;0)\right]^{L+1}} \; .
 \label{D=2f}
\end{align}
Define the operator \cite{SimmD, Loebb,Petkou0, Petkou, Petkou2}
\begin{equation}
\label{rd}
    R_d = \dfrac{1}{z - \bar{z}}\left(z\partial_z - \bar{z}\partial_{\bar{z}}\right),
\end{equation}
which was indicated in \cite{Loebb,Petkou0, Petkou, Petkou2}
as a shift operator in dimension of the space-time $D \to D+2$
for conformal (ladder) 4-point diagrams (see also~\cite[Theorem 49]{Borinsky:2021gkd}). Indeed, in our special
case we have the following statement.
\begin{proposition}\lb{pro1}
The function $\widetilde{\Phi}^{(\beta)}_L(z,\bar{z};\lambda)$, introduced in (\ref{beta0}), obeys the equation
 \be
 \lb{rd2}
 R_d \, \widetilde{\Phi}^{(\beta)}_L(z,\bar{z};\lambda) =
 \widetilde{\Phi}^{(\beta)}_L(z,\bar{z};\lambda+1) \; ,
 \ee
 i.e., since $\lambda = D/2-1$, the operator $R_d$ translates
 the expression for the conformal L-loop 4-point ladder integral for
 dimension $D$ to the expression for the conformal
 L-loop 4-point ladder integral for
 dimension $D+2$.
\end{proposition}
\noindent
{\bf Proof.} We prove the statement separately
 for the cases $\lambda >0$ and $\lambda=0$.
First, we consider the case $\lambda >0$.
Let us use the generating function of Gegenbauer polynomials
\begin{align}
\lb{derk}
\sum_{n = 0}^{\infty}
C^{(\lambda)}_n\left(r\right)\,t^n =
\frac{1}{(1-2r t+t^2)^{\lambda}}
\end{align}
to derive the differential
 equation for $C^{(\lambda)}_n(r)$. Indeed, we have
\begin{align}\label{derk02}
\partial_r \sum_{n = 0}^{\infty}
C^{(\lambda)}_n(r)\,t^n =
\frac{2\lambda t}{(1-2r t+t^2)^{\lambda+1}} =
2\lambda  \sum_{n = 0}^{\infty} C^{(\lambda+1)}_n(r) t^{n+1}
\;\;\; \Rightarrow
\end{align}
\be
\lb{derk03}
\partial_r  C^{(\lambda)}_n(r) = 2\lambda
C^{(\lambda+1)}_{n-1}(r) \; , \;\;\;\;\; \forall n \geq 1
 \; ; \;\;\;\;\;
\partial_r  C^{(\lambda)}_0(r) = 0 \; .
\ee

Let us apply the raising operator $R_d$ (\ref{rd})
to the right-hand side of (\ref{beta0}).
Due to $R_d\,(z\bar{z}) =0$, it acts nontrivially only
on the Gegenbauer polynomials, and in view of
(\ref{derk03}) the main formula is
\begin{align}\label{RC}
R_d\, C^{(\lambda)}_n\left(\frac{z+\bar{z}}{2\sqrt{z\bar{z}}}\right) = \frac{\lambda}{(z\bar{z})^{\frac{1}{2}}}\, C^{(\lambda+1)}_{n-1}\left(\frac{z+\bar{z}}{2\sqrt{z\bar{z}}}\right).
\end{align}
By means of this formula
 we obtain for \eqref{D=D}:
\begin{multline*}
R_d\, \widetilde{\Phi}^{(\beta)}_L(z ,\bar{z};\lambda)  =
\dfrac{\Gamma(\lambda)}{(z \bar{z})^{\lambda/2}} \,
\sum\limits_{n=0}^\infty
(n+\lambda)\left[ R_d\,
C^{(\lambda)}_n\left(\dfrac{z+\bar{z}}{2\sqrt{z\bar{z}}}\right)\right]
\int\limits_{-\infty}^{+\infty} d\nu
 \dfrac{(z\bar{z})^{{\sf i}\nu}}{\left[\tau_{n,\nu}(\beta;\lambda)\right]^{L+1}} =\\
\dfrac{\Gamma(\lambda)}{(z \bar{z})^{\lambda/2}} \,
\sum\limits_{n=1}^\infty
(n+\lambda)\frac{\lambda}{(z\bar{z})^{\frac{1}{2}}}\, C^{(\lambda+1)}_{n-1}\left(\frac{z+\bar{z}}{2\sqrt{z\bar{z}}}\right)
\int\limits_{-\infty}^{+\infty} d\nu
 \dfrac{(z\bar{z})^{{\sf i}\nu}}{\left[\tau_{n,\nu}(\beta;\lambda)\right]^{L+1}}  =  \\
\dfrac{\Gamma(\lambda+1)}{(z \bar{z})^{(\lambda+1)/2}} \,
\sum\limits_{n=0}^\infty
(n+1+\lambda) \, C^{(\lambda+1)}_{n}\left(\frac{z+\bar{z}}{2\sqrt{z\bar{z}}}\right)
\int\limits_{-\infty}^{+\infty} d\nu
 \dfrac{(z\bar{z})^{{\sf i}\nu}}{\left[\tau_{n,\nu}(\beta;\lambda+1)\right]^{L+1}}  =
 \widetilde{\Phi}^{(\beta)}_L(z ,\bar{z};\lambda+1).
\end{multline*}
where in the last line we shift $n \to n+1$ in the sum
and use the property
$\tau_{n+1,\nu}(\beta;\lambda)=\tau_{n,\nu}(\beta;\lambda+1)$
for the eigenvalues (\ref{tau1}).

Now we consider the second case $\lambda=0$
 (the transition $D=2 \to D=4$) and
 check that the function \eqref{D=2a}, \eqref{twodim-4d}
 satisfies \eqref{rd2} for $\lambda=0$:
 \be
\lb{cheb03}
 R_d  \widetilde{\Phi}^{(\beta)}_L(z ,\bar{z};0) \equiv
 \frac{1}{z - \bar{z}}(z \partial_z - \bar{z} \partial_{\bar{z}})
 \; \widetilde{\Phi}^{(\beta)}_L(z ,\bar{z};0) =
 \widetilde{\Phi}^{(\beta)}_L(z ,\bar{z};1) \; .
 \ee
  First, we note that after differentiating \eqref{cheb01}
  with respect to $r$, we
 deduce (see \cite{BatErd2},  section 10.11)
 \be
\lb{cheb04}
\partial_r T_n(r) = n C_{n-1}^{(1)}(r) \;\; \Rightarrow \;\;
\ee
$$
 \frac{1}{z - \bar{z}}(z \partial_z - \bar{z} \partial_{\bar{z}})
 T_n\Bigl(\frac{z +\bar{z}}{2\sqrt{z \bar{z}}} \Bigr) =
 \dfrac{ n }{2\sqrt{z \bar{z}}} C_{n-1}^{(1)}
 \Bigl(\frac{z +\bar{z}}{2\sqrt{z \bar{z}}} \Bigr)
$$
By making use of this relation
and identity $\tau_{n+1,\nu}(\beta;0)=
\tau_{n,\nu}(\beta;1)$, we find
for the action of $ R_d$ on \eqref{twodim-4d}, \eqref{D=2a}:
\be
\lb{cheb05}
\begin{array}{c}
\displaystyle
 R_d  \widetilde{\Phi}^{(\beta)}_L(z ,\bar{z};0)  =
 \frac{1}{\sqrt{z\bar{z}}} \sum\limits_{n=1}^\infty
   n\,  C_{n-1}^{(1)}
 \Bigl(\frac{z +\bar{z}}{2\sqrt{z \bar{z}}} \Bigr)
    \int\limits_{-\infty}^{+\infty} d\nu
     \dfrac{(z\bar{z})^{{\sf i}\nu}}{
     \left[\tau_{n,\nu}(\beta;0)\right]^{L+1}} = \\ [0.2cm]
     \displaystyle
     =  \frac{1}{\sqrt{z\bar{z}}} \sum\limits_{n=0}^\infty
   (n+1)\,  C_{n}^{(1)}
 \Bigl(\frac{z +\bar{z}}{2\sqrt{z \bar{z}}} \Bigr)
    \int\limits_{-\infty}^{+\infty} d\nu
     \dfrac{(z\bar{z})^{{\sf i}\nu}}{
     \left[\tau_{n,\nu}(\beta;1)\right]^{L+1}} =
     \widetilde{\Phi}^{(\beta)}_L(z ,\bar{z};1) \; ,
  \end{array}
\ee
where we take into account that the first term in the r.h.s.
of \eqref{twodim-4d}, \eqref{D=2a} is the zero mode
of the operator $R_d$. \hfill \qed

\subsection{Loop shift for generic propagator index $\beta$\label{sect:2.3}}
It is known (see~\cite{Loebb,Petkou0, Petkou, Petkou2}) that together with the dimensional shift operator~\eqref{rd} there exists an additional operator inducing recursion in the number of loops. Remarkably, this operator differs form the Laplace operator, which in the case of diagrams in Fig.~\ref{fig1}, also reduces the number of loops (for the reference see~\cite{DHSS, Drum, Borinsky:2021gkd, Borinsky:2022lds}). In this subsection we show that combining the operator $H_\beta$ with the dimensional shift operator, we can built an operator which increases the number of loops.

In our work \cite{DIS2}, we proved the following spectral relations:
\begin{align}
\label{H-op}
H_{\beta}\; | \psi^{\mu_1...\mu_n}_{\nu,\lambda} \rangle =
 \tau_{n,\nu}(\beta;\lambda) \;
 | \psi^{\mu_1...\mu_n}_{\nu,\lambda} \rangle \; ,
 \quad\quad
 \widehat{H}_{\beta}\;  \langle {\sf x}
 | \psi^{\mu_1...\mu_n}_{\nu,\lambda} \rangle =
  \tau_{n,\nu}(\beta;\lambda) \;
 \langle {\sf x} | \psi^{\mu_1...\mu_n}_{\nu,\lambda} \rangle,
\end{align}
where $\lambda=D/2-1$, $H_{\beta} = \hat{p}^{2\beta}\hat{q}^{2\beta}$,
$\widehat{H}_{\beta} = (-i \partial_{\sf x})^{2\beta}
\hat{\sf x}^{2\beta}$ and
 $$
 \langle {\sf x} | \psi^{\mu_1...\mu_n}_{\nu,\lambda} \rangle =
 \frac{{\sf x}^{\mu_1...\mu_n}}{(\sf x^2)^{\frac{\lambda +1+ n}{2}+\iu\nu}}
 \; , \quad \quad
 \tau_{n,\nu}(\beta;\lambda) =
 4^{\beta}\frac{\Gamma\left(\frac{\lambda + 1 + n}{2}+\beta-\iu\nu\right)
\Gamma\left(\frac{\lambda +1+ n}{2}+\iu\nu\right)}
{\Gamma\left(\frac{\lambda +1+ n}{2}-\beta+\iu\nu\right)
\Gamma\left(\frac{\lambda +1+ n}{2}-\iu\nu\right)} \; .
 $$
In the two-dimensional case ($\lambda$ = 0), the symmetric traceless tensor
${\sf x}^{\mu_1...\mu_n}$ has two components, which in
terms of complex coordinates (\ref{D2})
take the simple form $x^n$ and $\bar{x}^n$. Then,
we have ${\sf x}^2 = \bar{x} x$ and
using the complex coordinates write (\ref{H-op})
as two relations for $n\geq 0$
\begin{align}
\label{C4}
\widehat{H}_{\beta}\,\frac{1}{\sqrt{x\bar{x}}}\,x^{-\iu\nu+\frac{n}{2}}
\bar{x}^{-\iu\nu-\frac{n}{2}} &=\tau_{n\,,\nu}(\beta;0)\,
\frac{1}{\sqrt{x\bar{x}}}\,x^{-\iu\nu+\frac{n}{2}}
\bar{x}^{-\iu\nu-\frac{n}{2}} \ ;\ \\
\label{C5}
\widehat{H}_{\beta}\,\frac{1}{\sqrt{x\bar{x}}}\,x^{-\iu\nu-\frac{n}{2}}
\bar{x}^{-\iu\nu+\frac{n}{2}} &=\tau_{n\,,\nu}(\beta;0)\,
\frac{1}{\sqrt{x\bar{x}}}\,x^{-\iu\nu-\frac{n}{2}}
\bar{x}^{-\iu\nu+\frac{n}{2}} \; .
\end{align}
We substitute $x = z, \bar{x} = \bar{z}$ in the definition
of the two-dimensional operator $\widehat{H}_{\beta}$ 
and introduce the operator\footnote{Since $R_d$ has zero modes, the operator $(R_d)^{-\lambda}$ must be understood formally as an operator sending the function $\widetilde{\Phi}^{(\beta)}_{L}(z, \bar{z}; \lambda)$ to the function $\widetilde{\Phi}_{L}^{(\beta)}(z, \bar{z}; 0)$, the explicit form of which is known (see Proposition~\ref{pro-twodim-1} below).}
\begin{equation}
\label{rl}
    R^{(\beta)}_{\ell}(\lambda) := (R_d)^\lambda\sqrt{z\bar{z}}\,
    \widehat{H}_{-\beta}\,\dfrac{1}{\sqrt{z\bar{z}}}(R_d)^{-\lambda}.
\end{equation}
Now we formulate the following statement:
\begin{proposition}
The function $\widetilde{\Phi}_{L}^{(\beta)}(z,\bar{z}; \lambda)$ introduced in~\eqref{beta0} in the case $\lambda \in \mathbb{Z}_{>0}$ obeys the equations
\begin{equation}
\label{lshift}
    R_{\ell}^{(\beta)}(\lambda)\widetilde{\Phi}^{(\beta)}_L(z,\bar{z}; \lambda) = \widetilde{\Phi}_{L + 1}^{(\beta)}(z, \bar{z}; \lambda),
\end{equation}
where the operator $R_{\ell}^{(\beta)}(\lambda)$ is defined in~\eqref{rl}, i.e. it translates the expressions for the $L$-loop conformal $4$-point ladder integral to the $(L + 1)$-loop conformal $4$-point ladder integral.
\label{pro-Lshift}
\end{proposition}
\noindent
\textbf{Proof.}
First, we represent~\eqref{C4} and~\eqref{C5} for
$\beta \to - \beta$ and $x = z, \bar{x} = \bar{z}$ as
follows ($n \ge 0$):
\begin{align}
\label{C4z}
\widehat{H}_{-\beta}\,\frac{1}{\sqrt{z\bar{z}}}\,z^{-\iu\nu+\frac{n}{2}}
\bar{z}^{-\iu\nu-\frac{n}{2}} &=\tau_{n\,,\nu}(-\beta;0)\,
\frac{1}{\sqrt{z\bar{z}}}\,z^{-\iu\nu+\frac{n}{2}}
\bar{z}^{-\iu\nu-\frac{n}{2}} \ ;\ \\
\label{C5z}
\widehat{H}_{-\beta}\,\frac{1}{\sqrt{z\bar{z}}}\,z^{-\iu\nu-\frac{n}{2}}
\bar{z}^{-\iu\nu+\frac{n}{2}} &=\tau_{n\,,\nu}(-\beta;0)\,
\frac{1}{\sqrt{z\bar{z}}}\,z^{-\iu\nu-\frac{n}{2}}
\bar{z}^{-\iu\nu+\frac{n}{2}} \; .
\end{align}
By means of the symmetry (\ref{symtau}):
$\tau_{n\,,\nu}(\beta;0) = \tau_{-n\,,\nu}(\beta;0)$,
and after the change $\nu \to -\nu$,
we write (\ref{C4z}), (\ref{C5z})
 as one relation for $n\in \mathbb{Z}$
\begin{align}
\label{RL1}
\widehat{H}_{-\beta}\,
\frac{1}{\sqrt{z\bar{z}}}\,z^{\iu\nu+\frac{n}{2}}
\bar{z}^{\iu\nu-\frac{n}{2}} &= \tau_{n\,,-\nu}(-\beta;0) \,
\frac{1}{\sqrt{z\bar{z}}}\,z^{\iu\nu+\frac{n}{2}}
\bar{z}^{\iu\nu-\frac{n}{2}} \; , \\
\nonumber
\tau_{n\,,-\nu}(-\beta;0) \; &= \;
4^{-\beta}\frac{\Gamma\left(\frac{n+1}{2}-\beta+\iu\nu\right)
\Gamma\left(\frac{n+1}{2}-\iu\nu\right)}
{\Gamma\left(\frac{n+1}{2}+\beta-\iu\nu\right)
\Gamma\left(\frac{n+1}{2}+\iu\nu\right)}
\; \equiv \;  \bigl( \tau_{n\,,\nu}(\beta;0) \bigr)^{-1} \; .
\end{align}
Then we use formulas \eqref{rd2}, \eqref{RL1}
and representation \eqref{D=2f} to obtain
(for integer $\lambda$)
 \begin{align*}
 \widetilde{\Phi}_{L}^{(\beta)}(z, \bar{z}; \lambda) &=
(R_d)^{\lambda}\, \widetilde{\Phi}_{L}^{(\beta)}(z, \bar{z}; 0) =
 (R_d)^{\lambda}\,
\sum_{n \in \mathbb{Z}}\int\limits_{-\infty}^{\infty} d\nu\,
\bigl( \tau_{n\,,\nu}(\beta;0) \bigr)^{-(L + 1)}
z^{\iu\nu + \frac{n}{2}}\bar{z}^{\iu\nu - \frac{n}{2}} \\
 & =  (R_d)^{\lambda}\,\sqrt{z\bar{z}}\,
\widehat{H}_{-\beta}\,\frac{1}{\sqrt{z\bar{z}}}\,
\sum_{n \in \mathbb{Z}}\int\limits_{-\infty}^{\infty} d\nu\,
\bigl( \tau_{n\,,\nu}(\beta;0) \bigr)^{-L} \, z^{\iu\nu + \frac{n}{2}}\bar{z}^{\iu\nu - \frac{n}{2}} \\
& =  \left((R_d)^{\lambda}\,\sqrt{z\bar{z}}\,\widehat{H}_{-\beta}\,
\frac{1}{\sqrt{z\bar{z}}}\,
(R_d)^{-\lambda}\;
\right) \widetilde{\Phi}_{L-1}^{(\beta)}(z, \bar{z}; \lambda)
 \end{align*}
 Comparing the left-hand  and right-hand sides
 of this chain of relations for $L \to L+1$,
 we deduce (\ref{lshift}). \hfill\qed

Let us rewrite the operator
 $\widehat{H}_{-\beta} =
 (-i\partial_{\sf x})^{-2\beta} ({\sf x})^{-2\beta}$
 in the two-dimensional case in terms of the complex coordinates
 (see (\ref{D2}))
\begin{align}
({\sf x})^{2} = {\sf x}^2_1 +{\sf x}^2_2 = x\bar{x} \ \ ; \ \
(-i\partial_{\sf x})^{2}
= -\partial^2_{{\sf x}_1}-\partial^2_{{\sf x}_2} =
-4\,\partial_x\partial_{\bar{x}},
\end{align}
where we used the standard notation:
$\partial_x = \frac{1}{2}
\left(\partial_{{\sf x}_1}-\iu\partial_{{\sf x}_2}\right)$ and
$\partial_{\bar{x}} = \frac{1}{2}
\left(\partial_{{\sf x}_1}+\iu\partial_{{\sf x}_2}\right)$.
Then, after the change of
variables $x=z,\bar{x}=\bar{z}$ we have (for generic $\beta$)
\begin{align}
 \nonumber
\widehat{H}_{-\beta} =&
(-4\partial_z \partial_{\bar{z}})^{-\beta}\,
(z\bar{z})^{-\beta} = (-4)^{-\beta} (z^{\beta}\partial_z^\beta)^{-1} \,
(\bar{z}^\beta \partial_{\bar{z}}^{\beta})^{-1} \\
   = & (-4)^{-\beta} \,
 \frac{\Gamma(z \partial_z +1-\beta)
 \Gamma(\bar{z} \partial_{\bar{z}} +1-\beta)}{\Gamma(z \partial_z +1)
 \Gamma(\bar{z} \partial_{\bar{z}} +1)}
  \; .  \notag
\end{align}

 In order to analyze the operator $\big(R_{\ell}^{(\beta)}(\lambda)\big)^{-1}$,
 which shifts the loop number in the opposite
 direction $L \to L - 1$,
 we need the operator
\begin{align}
    \left(\widehat{H}_{-\beta}\right)^{-1} =& (-4)^{\beta}(z\bar{z})^{\beta}(\partial_{z}\partial_{\bar{z}})^{\beta} = (-4)^\beta(z^{\beta}\partial_z^\beta)(\bar{z}^\beta
    \partial_{\bar{z}}^{\beta}) \notag  \; .
\end{align}
For the four-dimensional case $\lambda = 1$
 (and generic $\beta$) the inverse of~\eqref{rl} is simplified
\begin{equation}
\label{RBL}
    \left(R^{(\beta)}_{\ell}(1)\right)^{-1} = (-4)^\beta R_d(z\bar{z})^{\beta + \frac{1}{2}}(\partial_{z}\partial_{\bar{z}})^{\beta}\dfrac{1}{(z\bar{z})^{\frac{1}{2}}}R_d^{-1} = (-4)^\beta\dfrac{(z\bar{z})^{\beta + \frac{1}{2}}}{z - \bar{z}}(\partial_z\partial_{\bar{z}})^{\beta}\dfrac{z - \bar{z}}{(z\bar{z})^{\frac{1}{2}}},
\end{equation}
where we used the explicit expression~\eqref{rd} for $R_d$. Here we also present the simplest example of the inverse relation to
relation~\eqref{lshift}
for $\beta=1$ and $D=4$ ($\lambda=1$):
\begin{align}
\label{C14}
-4\,\frac{\sqrt{z\bar{z}}}{z-\bar{z}}\,
z\partial_z\,\bar{z}\partial_{\bar{z}}\,
\frac{z-\bar{z}}{\sqrt{z\bar{z}}}\,\widetilde{\Phi}^{(1)}_L(z ,\bar{z};1) =
\widetilde{\Phi}^{(1)}_{L-1}(z ,\bar{z};1)
\end{align}
where we employed the inverse operator~\eqref{RBL} for $\beta=1$. This relation can be checked explicitly using the representation for the function $\widetilde{\Phi}^{(1)}_L(z, \bar{z}; 1)$ which we derive in subsection~\ref{sect:3.1}. We provide the corresponding calculations in Appendix~\ref{app:d}.

\noindent
{\bf Remark 6.} In \cite{Loebb}, the
authors demonstrated that the operator
(cf. (\ref{rd}))
\be
\lb{RL01}
\widetilde{R}_{\ell} = - \frac{1}{\log(z\bar{z})} \,
(z\partial_z + \bar{z}\partial_{\bar{z}}) \; ,
\ee
gives the recursion relations among conformal
ladder integrals for different
loop orders (for $L \to L-1$)
 in the case $\beta=1$ and arbitrary $\lambda=D/2-1$.
 Indeed, instead of
 (\ref{beta}) and (\ref{beta0}) for $\beta=1$,
we introduce the function
 \be
\lb{RL02}
\widetilde{\widetilde{\Phi}}^{(1)}_L(z,\bar{z};\lambda) =
L! \frac{(z \bar{z})^{\frac{\lambda-1}{2}}}{\Gamma(\lambda)} \,
\widetilde{\Phi}^{(1)}_L(z,\bar{z};\lambda)
\ee
and one can prove (see Appendix \ref{app:E})
that the function (\ref{RL02}) satisfies $\widetilde{R}_{\ell}
\widetilde{\widetilde{\Phi}}^{(1)}_L(z,\bar{z};\lambda) =
\widetilde{\widetilde{\Phi}}^{(1)}_{L-1}(z,\bar{z};\lambda)$
for generic $\lambda$.
Respectively, for the function $\widetilde{\Phi}^{(1)}_L$
 we obtain
\be
\lb{RL03}
 (z \bar{z})^{\frac{1-\lambda}{2}} \, 
 \widetilde{R}_{\ell} \, (z \bar{z})^{\frac{\lambda-1}{2}}
\widetilde{\Phi}^{(1)}_L(z,\bar{z};\lambda) = \frac{1}{L} \,
\widetilde{\Phi}^{(1)}_{L-1}(z,\bar{z};\lambda) \; .
\ee
 This relation for $\lambda=1$ and relation (\ref{C14})
give us the linear
 differential equation for the function
$\widetilde{\Phi}^{(1)}_{L}(z,\bar{z};1)$:
\be
\lb{RL04}
L \, \frac{1}{\log(z\bar{z})} \,
(z\partial_z + \bar{z}\partial_{\bar{z}})
\widetilde{\Phi}^{(1)}_{L}(z,\bar{z};1) =
4\,\frac{\sqrt{z\bar{z}}}{z-\bar{z}}\,
z\partial_z\,\bar{z}\partial_{\bar{z}}\,
\frac{z-\bar{z}}{\sqrt{z\bar{z}}}\,
\widetilde{\Phi}^{(1)}_{L}(z ,\bar{z};1) \; .
\ee
 Since the operator $(z\partial_z + \bar{z}\partial_{\bar{z}})$
 commutes with $\frac{(z-\bar{z})}{\sqrt{z\bar{z}}}$,
 the equation (\ref{RL04}) simplifies for the
 function $\Psi_L(z ,\bar{z}) : = \frac{z-\bar{z}}{\sqrt{z\bar{z}}}\,
\widetilde{\Phi}^{(1)}_{L}(z ,\bar{z};1)$
(see (\ref{der-D4}) below for explicit form of
this function):
\be
\lb{RL05}
\frac{L}{\log(z\bar{z})} \,
(z\partial_z + \bar{z}\partial_{\bar{z}})
\Psi_L(z ,\bar{z})  =
4\, z\partial_z\,\bar{z}\partial_{\bar{z}}\,
\Psi_L(z ,\bar{z})  \; .
\ee

\vspace{0.2cm}

\noindent
\textbf{Remark 7.} Another way of inducing the loop recursions relies on the application of the effective two-dimensional Laplace operator. For example, the authors of~\cite{Borinsky:2022lds} considered $D$-dimensional ladder diagrams by introducing a function which, in the case $\lambda = 1$, is connected with $\widetilde{\Phi}^{(1)}_{L}(z, \bar{z};1)$ by the rule (see~\cite[eq. (3.6)]{Borinsky:2022lds})
\begin{equation}
\label{Schnetz-funcs}
    \widetilde{f}_{L}(z, \bar{z}) = \dfrac{2^{2L + 1}}{\pi\sqrt{z\bar{z}}}\widetilde{\Phi}^{(1)}_{L}(z, \bar{z}; 1).
\end{equation}
The step of recursion then consists in using relation~\cite[eq. (3.2)]{Borinsky:2022lds} together with the following amputation of the line connecting $z$ and $0$ (see figure in~\cite[eq. (3.5)]{Borinsky:2022lds}), which results in
\begin{equation}
    \widetilde{f}_{L - 1}(z, \bar{z}) = z\bar{z}\cdot\left(-\dfrac{1}{z - \bar{z}}\,\partial_{z}\partial_{\bar{z}}\, (z - \bar{z})\right)\widetilde{f}_{L}(z, \bar{z}).
\end{equation}
Using rescaling~\eqref{Schnetz-funcs} in this relation, one immediately obtains~\eqref{C14}.
\section{Ladder integrals for even dimensions and integer $\beta$
\label{sect:3}}

\vspace{0.2cm}

It is desirable to evaluate the conformal integrals~\eqref{br1} and~\eqref{genf07}, corresponding to the $L$-loop ladder diagrams in Fig.~\ref{fig1} for an arbitrary choice of the parameter $\beta$ (assuming fixing the indices of lines as in~\eqref{genf00}, namely, index $\beta$ on the horizontal lines and $D/2 - \beta$ on the vertical lines). The one-integral representation~\eqref{beta} provides such a possibility, since the integration over $\nu$ can be performed by evaluating residues. The analytical result, however, seems to be feasible only for $\beta \in \mathbb{Z}_{>0}$ due to the significant simplifications in the pole structure.
\vspace{0.2cm}

In the following subsection~\ref{sect:3.1}, we argue that the one-integral representation~\eqref{beta} is sufficient to obtain an explicit analytical result in the case of even $D$ and $\beta = 1$. We perform such a derivation and show that the corresponding representation with Gegenbauer polynomials
\eqref{beta} for $\beta = 1$ can be systematically rewritten in the form that consists of classical polylogarithms with the coefficients being rational functions in $z$ and $\bar{z}$. The pole structure of~\eqref{beta} in the case of~$\beta = 2, 3\dots$ for $D \ge 4$ is quite involved, so instead of direct residue calculations we use the approach based on dimensional shift identities. Thus, in subsection~\ref{sect:fact}, we discuss the remarkable factorization occurring in the two-dimensional answer which, with the help of~\eqref{rd}, can be translated to any even dimension (note that this result holds for arbitrary $\beta$). In subsection~\ref{sect:3.2}, we then use this result for the case of $\beta = 2,3, \ldots$. Of particular interest is the mechanism of regularization of infrared singularities arising for the special combinations of $D$ and $\beta$.

\subsection{{Ladder integrals
and polylogarithms for $\beta=1$}\label{sect:3.1}}

For integer $\beta = k \in \mathbb{Z}_{>0}$ we have
(\ref{genf03a}) and the function (\ref{beta}) is written as
 \be
\lb{genf06}
\begin{array}{l}
\displaystyle
\Phi^{(\beta)}_L({\sf x},{\sf y})  
=  \dfrac{\Gamma(\lambda) \, L! \, 4^L  {\sf x}^{2\beta}}{
2 \pi^{\lambda+2}({\sf x}^2\, {\sf y}^2)^{D/4}} \,
\sum\limits_{n=0}^\infty
(n+\lambda)\, C^{(\lambda)}_n
 \left(\hat{{\sf x}}\hat{{\sf y}}\right)\, \; \times
 \\ [0.3cm]
\displaystyle
\quad\quad\quad\quad\quad\quad
 \times  \; \int\limits_{-\infty}^{+\infty} d\nu
 \, \frac{({\sf y}^2/{\sf x}^2)^{{\sf i}\nu}}{
 \Bigl(4^\beta \prod\limits_{m=1}^{\beta}
 (\frac{D}{4}+ \frac{n}{2}+{\sf i}\nu - m)
 (\frac{D}{4}+ \frac{n}{2}-1+m - {\sf i}\nu)\Bigr)^{L+1}} \; ,
 \end{array}
\ee
where $\lambda =D/2-1$. Further,
in this subsection, we consider the simplest case
$\beta = \alpha = 1$ ($\gamma=D/2-1$). This choice of parameters
in the case $D = 4$ leads to usual ladder diagrams with all indices on the lines equal to $1$. For arbitrary dimension $D$ the
 equation \eqref{genf06} represents the ladder diagram in Fig.~\ref{fig1} with indices $1$ on the horizontal lines and $D/2 - 1$ on the vertical lines. The corresponding integrals are convergent for any $D$, and we deduce from
 eq.~\eqref{genf06} the expression
\be
\lb{genf10}
\Phi^{(1)}_L({\sf x},{\sf y}) =  \dfrac{\Gamma(\lambda) \,
L! \;  {\sf x}^{2}}{
8 \pi^{\lambda+2}\; ({\sf x}^2\, {\sf y}^2)^{D/4}} \,
\sum\limits_{n=0}^\infty
(n+\lambda)\, C^{(\lambda)}_n
 \left(\hat{{\sf x}}\hat{{\sf y}}\right)
\int\limits_{-\infty}^{+\infty}
  \frac{d\nu \;\; ({\sf y}^2/{\sf x}^2)^{{\sf i}\nu}}{
 \Bigl((\frac{D}{4}+\frac{n}{2}-{\sf i}\nu)
 (\frac{D}{4}+\frac{n}{2}+{\sf i}\nu -1)\Bigr)^{L+1}} \,.
\ee
Integrating over $\nu$,
 we obtain the form\footnote{We close the contour of integration in the lower half-plane, assuming ${\sf x}^2 > {\sf y}^2$. The case of ${\sf x}^2 < {\sf y}^2$ is automatically taken into account due to the symmetry~\eqref{genf04d}.} (see
 eq. (3.47) in \cite{DIS2}\footnote{In \cite{DIS2},
 we used the notation $u,w$ for the vectors ${\sf x},{\sf y}$.})
\begin{multline}
\label{genf12}
\Phi^{(1)}_L({\sf x}, {\sf y}) = \dfrac{\Gamma(\lambda)}{
4\pi^{\lambda+1}\,({\sf x}^2)^{\lambda}}
\sum_{n = 0}^{\infty}\dfrac{C^{(\lambda)}_n
\left(\hat{{\sf x}}\hat{{\sf y}}\right)}{
({\sf x}^2/{\sf y}^2)^{n/2}}
\sum_{k = 0}^{L}\dfrac{(2L - k)!}{k!(L -k)!}
\dfrac{\log^k({\sf x}^2/{\sf y}^2)}{
(\lambda + n )^{2L - k}} = \\
\dfrac{\Gamma(\lambda)}{
4\pi^{\lambda+1}\,({\sf x}^2)^{\lambda}} \,
\sum_{k = 0}^{L}\dfrac{(2L - k)!}{k!(L -k)!}\,
\log^k({\sf x}^2/{\sf y}^2)\,
\sum_{n = 0}^{\infty}\dfrac{C^{(\lambda)}_n
\left(\hat{{\sf x}}\hat{{\sf y}}\right)}{
({\sf x}^2/{\sf y}^2)^{n/2}}\,
\dfrac{1}{(\lambda + n)^{2L - k}}.
\end{multline}
Then, the function~\eqref{genf12} can be written
(with the help of the parametrization (\ref{genf11})) as
\begin{equation}
\lb{der00}
    \Phi^{(1)}_L({\sf x}, {\sf y}) =
    \dfrac{\Gamma(\lambda)}{4\pi^{\lambda+ 1}{\sf x}^{2\lambda}}\sum_{k = 0}^{L}\dfrac{(-1)^k(2L - k)!}{k!(L - k)!}\log^k(z\bar{z})\Sigma^{(\lambda)}_s(z,\bar{z}),
\end{equation}
where we introduce
 \be
 \lb{der01}
 \Sigma_{s}^{(\lambda)}(z, \bar{z}) = \sum_{n = 0}^{\infty}\dfrac{C^{(D/2-1)}_n\left(\hat{{\sf x}}\hat{{\sf y}}\right)}{
({\sf x}^2/{\sf y}^2)^{n/2}}\,
\dfrac{1}{(D/2 + n - 1)^{s}} = \sum_{n = 0}^{\infty}
C^{(\lambda)}_n\left(\frac{z+\bar{z}}{2\sqrt{z\bar{z}}}\right)
(z\bar{z})^{n/2}\,
\dfrac{1}{(\lambda + n)^{s}} \; ,
\ee
and denote $s= 2L - k$. In what follows, we show that~\eqref{der01} can be expressed in terms of known special functions
(in particular, classical polylogarithms and rational functions of $z$ and $\bar{z}$).
\begin{proposition}
\label{test1}
 The function (\ref{der01}) can be expressed in the following form:
 \be
 \lb{der02}
 \Sigma_{s}^{(\lambda)}(z, \bar{z})  =
 P_{\lambda}(z\partial_z)\left(\frac{z^\lambda}{(z-\bar{z})^\lambda}
 \Phi(z,s,\lambda)\right) +
 P_{\lambda}(\bar{z}\partial_{\bar{z}}) \left(
 \frac{\bar{z}^\lambda}{(\bar{z}-z)^\lambda}
 \Phi(\bar{z},s,\lambda) \right),
\ee
 where
  \be
 \lb{der04}
 P_{\lambda}(z\partial_z) =
 \frac{1}{\Gamma(\lambda)} \frac{\Gamma(z\partial_z +\lambda)}{\Gamma(z\partial_z +1)} \; ,
 \ee
 and
   \be
 \lb{der04a}
 \Phi(z,s,\lambda) = \sum_{n=0}^\infty \frac{z^n}{(\lambda + n)^s}
 \ee
 is the Lerch function (see \cite{BatErd}), which generalizes the
 polylogarithms
  $$
  \Li_s(z)= z \, \Phi(z,s,1) =
  \sum_{n=1}^\infty \frac{z^n}{n^s} \; .
  $$
  For convergence
 of the infinite sum in (\ref{der04a}) we require
 $\lambda \in \mathbb{R}_{> 0}$, $|z| < 1$.
  \end{proposition}
  \noindent
  {\bf Proof.}
  First, we rewrite our expression
(\ref{der01}) in the following form:
\begin{multline}
\label{der06}
\Sigma_{s}^{(\lambda)}(z, \bar{z}) =
\left.\dfrac{1}{(\lambda+x\partial_x)^{s}}\,
\sum_{n = 0}^{\infty}
C^{(\lambda)}_n\left(\frac{z+\bar{z}}{2\sqrt{z\bar{z}}}\right)
(z\bar{z})^{n/2}\,x^n \right|_{x=1} = \\
= \left.\dfrac{1}{(\lambda+x\partial_x)^{s}}
\frac{1}{\left(1-(z+\bar{z})x+z\bar{z}\,x^2\right)^{\lambda}}\right|_{x=1} =
\left.\dfrac{1}{(\lambda+x\partial_x)^{s}}
\frac{1}{\left(1-z x\right)^{\lambda}\left(1-\bar{z} x\right)^{\lambda}}\right|_{x=1},
\end{multline}
where we apply an obvious identity
$\left. f(x\partial_x) \,x^n \right|_{x=1} = f(n)$
and substitute the explicit expression
(\ref{derk}) of the generating function of Gegenbauer polynomials
for $t = x \sqrt{z\bar{z}}$ and $r = \frac{z+\bar{z}}{2\sqrt{z\bar{z}}}$.
Then, we note that
 \be
 \lb{der03}
 \frac{1}{\left(1-z x\right)^{\lambda}} =
 P_\lambda(z\partial_z) \cdot \frac{1}{\left(1-z x\right)}  \; ,
 \ee
 where the operator $P_\lambda(z\partial_z)$ was introduced in (\ref{der04}).
Indeed, we have
\begin{multline*}
\frac{1}{\left(1-z x\right)^{\lambda}} =
\sum_{n=0}^{\infty} \frac{\Gamma(n+\lambda)}{n! \Gamma(\lambda)} z^n x^n =
\frac{1}{\Gamma(\lambda)}\,\sum_{n=0}^{\infty} \frac{\Gamma(z\partial_z+\lambda)}{\Gamma(z\partial_z+1)} \, z^n x^n =
\frac{\Gamma(z\partial_z+\lambda)}{\Gamma(\lambda) \,\Gamma(z\partial_z+1)} \cdot
\frac{1}{1-z x}.
\end{multline*}
The substitution of (\ref{der03})
(and its analog for $z \to \bar{z}$) into (\ref{der06}) gives
\be
\lb{der07}
\Sigma_{s}^{(\lambda)}(z, \bar{z}) =
\left.\dfrac{1}{(\lambda+x\partial_x)^{s}}
P_\lambda(z\partial_z) P_\lambda(\bar{z}\partial_{\bar{z}})
\frac{1}{\left(1-z x\right) \, \left(1-\bar{z} x\right)}\right|_{x=1}.
\ee

At the next step, we perform the partial fraction decomposition in the r.h.s. of~\eqref{der07}
\begin{align}
\frac{1}{\left(1-z x\right)\left(1-\bar{z} x\right)} =
\frac{z}{z-\bar{z}}\frac{1}{\left(1-z x\right)} -
\frac{\bar{z}}{z-\bar{z}}\frac{1}{\left(1-\bar{z} x\right)}.
\end{align}
We use the definition of the Lerch function (\ref{der04a}):
\begin{equation}
\label{der08}
\left.\dfrac{1}{(\lambda+x\partial_x)^{s}}
\frac{1}{\left(1-z x\right)}\right|_{x=1} =
\sum_{n=0}^{\infty} \dfrac{z^n}{(\lambda+n)^{s}} \equiv \Phi(z,s,\lambda) \; ,
\end{equation}
and its conjugation counterpart for $z \leftrightarrow \bar{z}$.
 As a result,
from (\ref{der07}) we deduce the relation
\be
\lb{der09}
 \Sigma_{s}^{(\lambda)}(z, \bar{z})  =
 P_{\lambda}(z\partial_z)  P_{\lambda}(\bar{z}\partial_{\bar{z}})
 \left(\frac{z}{(z-\bar{z})} \Phi(z,s,\lambda) +
 \frac{\bar{z}}{(\bar{z}-z)}
 \Phi(\bar{z},s,\lambda) \right) \; .
\ee
Here we again use relation (\ref{der03}) for $x=\bar{z}^{-1}$
(and its analog for $z \leftrightarrow \bar{z}$)
and finally derive  (\ref{der02}). \hfill \qed

\vspace{0.3cm}

Note that the operator of dilatation $z\partial_z$ applied to the Lerch function shifts its parameter, namely,
\begin{align}
\label{der13}
(z\partial_z  + \lambda) \Phi(z,s,\lambda) = \Phi(z,s-1,\lambda).
\end{align}
Thus, the application of the differential operator
in (\ref{der02}) can be reduced to
simple transformations of the Lerch functions $\Phi(z,s,\lambda)$.
In what follows we will limit ourselves to the case of positive integer $\lambda = D/2-1 \in \mathbb{Z}_{>0}$,
which corresponds to the even dimensions $D=2(1+\lambda)$.
 For such a choice of $\lambda$,
 the operator $P_{\lambda}(z\partial_z)$ in (\ref{der10})
 becomes a polynomial in $z\partial_z$:
  \be
 \lb{der10}
 P_{\lambda}(z\partial_z) =
 \frac{1}{\Gamma(\lambda)} \frac{\Gamma(z\partial_z +\lambda)}{\Gamma(z\partial_z +1)} =
 \frac{1}{(\lambda-1)!}
 (z\partial_z+1)\cdots(z\partial_z+\lambda-1) \equiv
 \frac{1}{\Gamma(\lambda)} \,
\partial_z^{\lambda-1} \, z^{\lambda-1}  \; ,
\ee
with $P_{1}(z\partial_z) =1$ and one can directly apply
formula (\ref{der13}) in (\ref{der02}),
(\ref{der09}) to obtain an
 explicit expression for $\Sigma_{s}^{(\lambda)}(z, \bar{z})$.
 We note that the representation in the right-hand side of (\ref{der10}), which follows form the fact $\big[\partial_z, z^p\big] = pz^{p - 1}$,
 is more convenient for the analytical continuation
 of  $P_{\lambda}(z\partial_z)$ for $\lambda \in \mathbb{C}$.

\begin{proposition}\lb{pro2}
 In the case $\lambda = D/2-1 \in \mathbb{Z}_{>0}$,
 the function $\Sigma_{s}^{(\lambda)}(z, \bar{z})$
  defined in (\ref{der02}) has an explicit
 expression in terms of polylogarithms
 \be
 \lb{der16}
 \Sigma_s^{(\lambda)}(z, \bar{z}) = P_{\lambda}(z\partial_z) \frac{\Li_s(z)}{(z-\bar{z})^{\lambda}} +
 P_{\lambda}(\bar{z}\partial_{\bar{z}})
 \frac{\Li_s(\bar{z})}{(\bar{z}-z)^{\lambda}},
 \ee
 where the operators $P_{\lambda}(z\partial_z)$
 and $P_{\lambda}(\bar{z}\partial_{\bar{z}})$ were
 defined in (\ref{der10}).
\end{proposition}
{\bf Proof.} For $\lambda \in \mathbb{Z}_{>0}$, the
Lerch function (\ref{der04a}) is reduced to the
sum of a polylogarithm and a
polynomial in $z^{-1}$.
Indeed, for $\lambda > 1$ we have
\be
 \lb{der12}
 \Phi(z,s,\lambda) = \frac{1}{z^\lambda} \sum_{n = 0}^{\infty}
 \dfrac{z^{\lambda+n}}{(\lambda+n)^s} =
 \frac{1}{z^\lambda} \sum_{n = \lambda}^{\infty}
 \dfrac{z^{n}}{n^s}  = \frac{1}{z^\lambda} \left(\Li_s(z)  - \sum_{n = 1}^{\lambda-1}\dfrac{z^{n}}{n^s}\right) \, .
 \ee
Applying this relation to~\eqref{der09}, we get
 \be
\lb{der14}
\begin{array}{c}
 \Sigma_{s}^{(\lambda)}(z, \bar{z})  =
 P_{\lambda}(z\partial_z)  P_{\lambda}(\bar{z}\partial_{\bar{z}})
 \left(\frac{z^{1-\lambda}}{z-\bar{z}} \Bigl(\Li_s(z)  -
 \sum\limits_{n = 1}^{\lambda-1}\dfrac{z^{n}}{n^s}\Bigr) +
 \left(z \leftrightarrow \bar{z}\right) \right) = \\ [0.3cm]
 =
 P_{\lambda}(z\partial_z)  P_{\lambda}(\bar{z}\partial_{\bar{z}})
 \left(\Bigl(\frac{z^{1-\lambda}}{z-\bar{z}} \Li_s(z)  +
 \left(z \leftrightarrow \bar{z}\right) \Bigr)
  - \sum\limits_{n = 1}^{\lambda-1}\dfrac{(z^{n+1-\lambda}-
 \bar{z}^{n+1-\lambda})}{n^s(z-\bar{z})}
  \right) = \\ [0.3cm]
 = P_{\lambda}(z\partial_z)P_{\lambda}(\bar{z}\partial_{\bar{z}})
 \left(\frac{z^{1-\lambda}}{z-\bar{z}} \Li_s(z)  +
 \frac{\bar{z}^{1-\lambda}}{\bar{z}-z} \Li_s(\bar{z}) \right)
\; .
 \end{array}
\ee
Here we use the identity\footnote{For $\lambda =1,2$
the expression in the left hand-side
 of (\ref{der15}) automatically vanishes.} $(\lambda > 2)$
\be
\lb{der15}
\begin{array}{c}
P_{\lambda}(z\partial_z)P_{\lambda}(\bar{z}\partial_{\bar{z}})
\sum\limits_{n = 1}^{\lambda-1}\dfrac{(z^{n+1-\lambda}-
 \bar{z}^{n+1-\lambda})}{n^s(z-\bar{z})} = \\ [0.3cm]
 = P_{\lambda}(z\partial_z)P_{\lambda}(\bar{z}\partial_{\bar{z}})
\sum\limits_{n = 1}^{\lambda-2}\frac{(- 1)}{n^s}
\sum\limits_{m = 1}^{\lambda-n-1} z^{-m}\,
 \bar{z}^{\, n+m-\lambda} = 0 \; ,
 \end{array}
\ee
which follows from the fact that the operator $P_{\lambda}(z\partial_z)$
defined in (\ref{der10})
and its analog $P_{\lambda}(\bar{z}\partial_{\bar{z}})$
have zero modes $z^{-1}\,,z^{-2}\,,...\,,z^{1-\lambda}$
and $\bar{z}^{-1}\,,\bar{z}^{-2}\,,...\,,\bar{z}^{1-\lambda}$.
Thus, the sum over $m$
appearing in (\ref{der15})
evidently belongs to the kernel of the product
of these operators.

Finally, we apply formula (\ref{der03})
 and its analog for $z \to \bar{z}$ to equation (\ref{der14})
to obtain (\ref{der16}).
\hfill \qed

We note that
the substitution of (\ref{der14})
into (\ref{der00}) reproduces
(up to normalization)
 the result~\cite[eq. (3.6)]{Borinsky:2022lds}.

\vspace{0.3cm}

The analog of
proposition {\sf \ref{pro1}} exists for the function $\Sigma_{s}^{(\lambda)}(z, \bar{z})$.
Indeed, we have
\begin{proposition}
\lb{pro2b}
The functions (\ref{der16}) satisfy the relation
\be
\lb{rd01}
    \frac{1}{\lambda} R_d \cdot \Sigma_{s}^{(\lambda)}(z, \bar{z}) =
   \Sigma_{s}^{(\lambda+1)}(z, \bar{z}) \;  ,
\ee
 where the recursion operator $R_d$ was defined in (\ref{rd}).
\end{proposition}
\noindent
{\bf Proof.} Since the function $\Sigma_{s}^{(\lambda)}(z, \bar{z})$
in (\ref{der16}) is represented as the sum
$$
\Sigma_s^{(\lambda)}(z, \bar{z}) =  \Sigma_s^{\prime (\lambda)}(z, \bar{z}) +
\Sigma_s^{\prime (\lambda)}(\bar{z},z) \; , \;\;\;\;\;\;
\Sigma_s^{\prime (\lambda)}(z, \bar{z}) : =
\frac{1}{\Gamma(\lambda)} \,
\partial_z^{\lambda-1} \, z^{\lambda-1}\,
\frac{1}{(z-\bar{z})^\lambda} \Li_s(z)
$$
and in view of the symmetry
$R_d|_{z \leftrightarrow \bar{z}} = R_d$,
we need only to prove the identity
\be
\lb{rd02}
    \frac{1}{\lambda} R_d \cdot
    \Sigma_{s}^{\prime (\lambda)}(z, \bar{z}) =
   \Sigma_{s}^{\prime (\lambda+1)}(z, \bar{z}) \;  .
\ee
Indeed, we have
$$
\begin{array}{c}
\frac{1}{\lambda} R_d \cdot
    \Sigma_{s}^{\prime (\lambda)}(z, \bar{z}) =
\frac{1}{\lambda} \dfrac{1}{z - \bar{z}}(z\partial_z - \bar{z}\partial_{\bar{z}}) \frac{1}{\Gamma(\lambda)} \,
\partial_z^{\lambda-1} \, z^{\lambda-1}
\frac{1}{(z-\bar{z})^\lambda} \Li_s(z) = \\ [0.3cm]
= \frac{1}{\Gamma(\lambda+1)} \, \dfrac{1}{z - \bar{z}}
\partial_z^{\lambda-1} \, z^{\lambda-1}
(z\partial_z - \bar{z}\partial_{\bar{z}})
\frac{1}{(z-\bar{z})^\lambda} \Li_s(z) = \\ [0.3cm]
= \frac{1}{\Gamma(\lambda+1)} \, \dfrac{1}{z - \bar{z}}
\partial_z^{\lambda-1} \, z^{\lambda-1} \left(
z\partial_z \, \frac{1}{(z-\bar{z})^\lambda}
 -
\frac{\lambda \bar{z}}{(z-\bar{z})^{\lambda+1}} \right) \Li_s(z) =
\\ [0.3cm]
= \frac{1}{\Gamma(\lambda+1)} \, \dfrac{1}{z - \bar{z}}
 \partial_z^{\lambda-1} \, \left(  z^{\lambda}
\partial_z  -
\lambda  \, z^{\lambda-1}
\dfrac{\bar{z}}{(z-\bar{z})} \right)
\, \dfrac{1}{(z-\bar{z})^\lambda} \Li_s(z) =
\\ [0.3cm]
= \frac{1}{\Gamma(\lambda+1)} \, \dfrac{1}{z - \bar{z}}
\left( \partial_z^{\lambda}  -
\lambda \partial_z^{\lambda-1} \dfrac{1}{(z-\bar{z})}
 \right) \dfrac{z^\lambda}{(z-\bar{z})^\lambda} \Li_s(z)
= \frac{1}{\Gamma(\lambda+1)} \, \partial_z^{\lambda}
 \dfrac{z^\lambda}{(z-\bar{z})^{\lambda+1}} \Li_s(z).
\end{array}
$$
In the last equality we use the identity
$$
\dfrac{1}{z - \bar{z}}
\left( \partial_z^{\lambda}  -
\lambda \partial_z^{\lambda-1} \dfrac{1}{(z-\bar{z})}
 \right) = \partial_z^{\lambda} \dfrac{1}{z - \bar{z}},
$$
which follows from the obvious relation
$\partial_z^{\lambda} (z - \bar{z}) =
(z - \bar{z})\partial_z^{\lambda} +
\lambda \partial_z^{\lambda-1}$.

\hfill \qed

\vspace{0.5cm}

We argue that the representation of the function $\Sigma_s^{(\lambda)}(z,\bar{z})$ in the form~\eqref{der16} is already sufficient to build a recurrent procedure for obtaining explicit expressions for any $\lambda \in \mathbb{Z}_{\ge 0}$. In order to proceed with such calculations, it might be useful to note the property of the dilatation operator $z \partial_z$;
 its application to the polylogarithm shifts its weight
    \begin{equation}
    \label{shu02}
        z\partial_z\Li_s(z) = \Li_{s - 1}(z),
    \end{equation}
    which follows directly from~\eqref{der13} in the case of $\lambda =1$.

\noindent
\textbf{Remark 8.} Note that despite the fact that relations~\eqref{rd2} and~\eqref{rd01} were derived from the different starting points,
namely, one from the general representation~\eqref{beta0} for generic $\beta$ and the other from the application of the differential operator $P_{\lambda}(z\partial_z)$ in~\eqref{der16}, we can easily see their connection. THe operator $R_d$ commutes with any function
$f(z\bar{z})$ and in particular with $\log(z\bar{z})$,
\begin{equation*}
    \big[R_d, \log(z\bar{z})\big] = 0,
\end{equation*}
which is needed
 to shift the application of this operator from $\Sigma_s^{(\lambda)}(z, \bar{z})$ to the whole function $\tilde{\Phi}_L^{(\lambda)}(z,\bar{z}; \lambda)$.

\vspace{0.3cm}

\noindent
{\bf Remark 9.} In view of relation (\ref{der00}),
the function $\Phi_L({\sf x}, {\sf y})$ depends
only on the product
$\widetilde{\Sigma}_s^{(\lambda)} = \Gamma(\lambda) \Sigma_s^{(\lambda)}$. According to~\eqref{rd01},
the function $\widetilde{\Sigma}_s^{(\lambda)}$
satisfies the equation
\be
\lb{rd05}
 R_d \cdot \widetilde{\Sigma}_s^{(\lambda)}
 (z, \bar{z}) =
   \widetilde{\Sigma}_s^{(\lambda+1)}(z, \bar{z}) \;  .
\ee
Using this relation, we can  define the two-dimensional function
\begin{equation}
\label{twodim-sigma}
    \widetilde{\Sigma}_s^{(0)} = \Li_{s + 1}(z) + \Li_{s + 1}(\bar{z}) + X(z\bar{z}),
\end{equation}
where $X(z\bar{z})$ is the zero mode of the operator $R_d$. One can see that the two-dimensional function~\eqref{D=2a} is singular in the case $\beta = 1$, which means that in order to relate~\eqref{D=2a} and~\eqref{twodim-sigma} $X(z\bar{z})$ should correspond to a singular contribution. For more details, see subsection~\ref{sect:3.2}.


\vspace{0.2cm}
\noindent
{\bf Examples.} Here we present the explicit answers for the~\eqref{der16} in terms of polylogaritms and rational functions for several small values of $\lambda$.
\begin{itemize}
    \item $\lambda = 1$:
    \begin{equation*}
        \Sigma_s^{(1)}(z, \bar{z}) = \dfrac{\Li_s(z)}{z - \bar{z}} + (z \leftrightarrow \bar{z}).
    \end{equation*}
    \item $\lambda = 2$:
    \begin{align*}
        \Sigma^{(2)}_s(z, \bar{z}) =& (z\partial_z + 1)\dfrac{\Li_s(z)}{(z - \bar{z})^2} + (z \leftrightarrow \bar{z}) \notag \\ =& -\dfrac{z + \bar{z}}{(z - \bar{z})^3}\Li_s(z) + \dfrac{\Li_{s - 1}(z)}{(z - \bar{z})^2} + (z \leftrightarrow \bar{z}). %
    \end{align*}
    \item $\lambda = 3$:
    \begin{align*}
        \Sigma_s^{(3)}(z, \bar{z}) =&  \dfrac{1}{2}(z\partial_z + 2)(z\partial_z + 1)\dfrac{\Li_s(z)}{(z -\bar{z})^3} + (z \leftrightarrow \bar{z})\notag \\ & = \dfrac{z^2 + 4z\bar{z} + \bar{z}^2}{(z - \bar{z})^5}\Li_{s}(z)-\dfrac{3(z + \bar{z})}{2(z - \bar{z})^4}\Li_{s - 1}(z) + \dfrac{\Li_{s - 2}(z)}{2(z - \bar{z})^3} + (z \leftrightarrow \bar{z}). 
    \end{align*}
    \item $\lambda = 4$:
    \begin{align*}
        \Sigma_s^{(4)}(z, \bar{z}) =& \dfrac{1}{6}(z\partial_z + 3)(z\partial_z + 2)(z\partial_z + 1)\dfrac{\Li_s(z)}{(z - \bar{z})^{4}} + (z \leftrightarrow \bar{z})\notag \\ =&
    -\dfrac{(z+\bar{z})(z^2 + 8z \bar{z} + \bar{z}^2)}{(z - \bar{z})^7}
        \Li_s(z) + \dfrac{11z^2 + 38z\bar{z} + 11\bar{z}^2}{6(z - \bar{z})^6}\Li_{s - 1}(z)\notag \\ & - \dfrac{z + \bar{z}}{(z - \bar{z})^5}\Li_{s - 2}(z) + \dfrac{\Li_{s - 3}(z)}{6(z - \bar{z})^4} + (z \leftrightarrow \bar{z}).
    \end{align*}
    \item  $\lambda =5$:
    \begin{align*}
        \Sigma_{s}^{(5)}(z, \bar{z}) =&
        \frac{1}{24}(z\partial_z + 4)(z\partial_z + 3)(z\partial_z + 2)(z\partial_z + 1)
        \frac{1}{(z-\bar{z})^5}\,\Li_s(z) +
        (z \leftrightarrow \bar{z})  \notag\\
         =&\frac{z^4+16z^3\bar{z}+
         36z^2\bar{z}^2+
         16z\bar{z}^3+\bar{z}^4}{(z-\bar{z})^9}\,\Li_{s}(z) -\frac{5(z+\bar{z})(5z^2 +32z\bar{z} +5\bar{z}^2)}{12(z-\bar{z})^{8}}\,\Li_{s-1}(z) \notag
          \\
         &+\frac{5(7z^2+22z\bar{z}+7\bar{z}^2)}{24(z-\bar{z})^7}\,\Li_{s-2}(z) -
         \frac{5(z+\bar{z})}{12(z-\bar{z})^6} \Li_{s-3}(z) + \dfrac{\Li_{s-4}(z)}{24(z- \bar{z})^5}  +
        (z \leftrightarrow \bar{z}).
    \end{align*}
\end{itemize}

\vspace{0.3cm}

Analyzing the examples for small values of $\lambda$, we can clearly see the pattern. For a given $\lambda$, the answer for $\Sigma^{(\lambda)}_s$
is expressed in terms of the number of polylogarithms $\Li_{s}(z), \Li_{s - 1}(z), \ldots, \Li_{s - \lambda + 1}(z)$ accompanied by rational functions of the form $\frac{G_k(z, \bar{z})}{(z - z)^{2\lambda - k - 1}}$, where $G_k(z, \bar{z})$ is a homogeneus polynomial of degree $\lambda - 1 - k$ symmetric under the transformation $z \leftrightarrow \bar{z}$. In Appendix~\ref{app:a}, we give a compact formula for the polynomial $G_k^{(\lambda)}(z, \bar{z})$ and show that properties such as homogeneity and symmetry hold for any $\lambda \in \mathbb{Z}_{>0}$.

 \vspace{0.3cm}

At the end of this subsection
we give
explicit expressions for the functions (\ref{der00})
and (\ref{beta0})$|_{\beta=1}$
 of $L$-loop ladder integrals in $D=4,6,8,10$. \\
{\bf 1.} The case $D=4$ ($\lambda=D/2-1=1$)
and indices on the lines $\alpha=\beta=\gamma=1$
\cite{Usyukina:1992jd,Usyukina:1993ch} (see also
\cite{Isa,DIS2}):
\begin{equation}
\lb{der-D4}
\begin{array}{c}
 \widetilde{\Phi}^{(1)}_L(z, \bar{z};1) =
    \dfrac{\pi \, (z \bar{z})^{1/2}}{2(z - \bar{z})4^L}
    \sum\limits_{k = 0}^{L}  
   C^L_{2L-k} \, \dfrac{(-1)^k\log^k(z\bar{z})}{k!}  
 \Bigl(\Li_{2L-k}(z)-\Li_{2L-k}( \bar{z})\Bigr) \, . 
 \end{array}
\end{equation}
{\bf 2.} The case $D=6$ ($\lambda=2$)
and indices on the lines $\alpha=\beta=1$,
$\gamma=2$~\cite{Loebb}:
\begin{equation}
\lb{der-D6a}
\begin{array}{l}
    \widetilde{\Phi}^{(1)}_L(z, \bar{z};2)  =
   \dfrac{\pi \, (z \bar{z})^{1/2}}{2(z - \bar{z})^2 \, 4^L}
    \sum\limits_{k = 0}^{L}
   C^L_{2L-k}  \, \dfrac{(-1)^k\log^k(z\bar{z})}{k!} 
 \;   \\ [0.3cm]
  \quad\quad \Bigl(-\dfrac{(z + \bar{z})}{(z - \bar{z})}
  (\Li_{2L-k}(z)-\Li_{2L-k}(\bar{z}) )+
  \Li_{2L-k-1}(z)+\Li_{2L-k-1}(\bar{z})
  \Bigr)\, .
  \end{array}
\end{equation}
{\bf 3.} The case $D=8$ ($\lambda=3$)
and indices on the lines $\alpha=\beta=1$,
$\gamma=3$:
\begin{equation}
\lb{der-D8a}
\begin{array}{l}
\widetilde{\Phi}^{(1)}_L(z, \bar{z};3)  =
   \dfrac{\pi \, (z \bar{z})^{1/2}}{2(z - \bar{z})^3 \, 4^L}
    \sum\limits_{k = 0}^{L}
 C^L_{2L-k}  \, \dfrac{(-1)^k\log^k(z\bar{z})}{k!}  
    \;  \\ [0.4cm]
  \Bigl(2
  \dfrac{(z^2 +4z\bar{z}+ \bar{z}^2)}{(z - \bar{z})^2}
  \Li_{2L-k}(z)+ 3\dfrac{(z + \bar{z})}{(z - \bar{z})}
  \Li_{2L-k-1}(z)+ \Li_{2L-k-2}(z)
  + (z \leftrightarrow \bar{z}) \Bigr)\, .
  \end{array}
\end{equation}
{\bf 4.} The case $D=10$ ($\lambda=4$)
and indices on the lines $\alpha=\beta=1$,
$\gamma=4$:
\begin{equation}
\lb{der-D10a}
\begin{array}{l}
\widetilde{\Phi}^{(1)}_L(z, \bar{z};4)  =
\dfrac{\pi \, (z \bar{z})^{1/2}}{2(z - \bar{z})^4 \, 4^L}
    \sum\limits_{k = 0}^{L}
   C^L_{2L-k} \, \dfrac{(-1)^k\log^k(z\bar{z})}{k!}  
    \Bigl(
    -6\dfrac{(z+\bar{z})(z^2 +8z\bar{z}+\bar{z}^2)}{
    (z-\bar{z})^3}\Li_{2L-k}(z) + \\ [0.4cm]
 + \dfrac{(11z^2 +38z\bar{z}+ 11\bar{z}^2)}{(z - \bar{z})^2}  \Li_{2L-k-1}(z)-
  6\dfrac{(z+\bar{z})}{(z - \bar{z})}
  \Li_{2L-k-2}(z)+   \Li_{2L-k-3}(z)
  + (z \leftrightarrow \bar{z}) \Bigr)\, ,
  \end{array}
\end{equation}
where $C^L_{2L-k} = 
(2L-k)!/((L-k)!L!)$ -- are binomial coefficients.

\subsection{Two-dimensional factorization for generic $\beta$ \label{sect:fact}}
In subsection~\ref{sect:2.2} we concluded that the $D$-dimensional conformal ladder diagrams with the  generic index~$\beta$ for $D > 2$ and $D = 2$
are respectively presented in the form~\eqref{D=D}
and~\eqref{D=2a}.
 We also explicitly showed that the representation~\eqref{D=D} admits a dimensional shift with the help of the operator~\eqref{rd}~\cite{Loebb, Petkou0, Petkou, Petkou2}, namely
\begin{equation}
\label{twodim-5}
    R_d \, \widetilde{\Phi}^{(\beta)}_L(z,\bar{z};\lambda) =
 \widetilde{\Phi}^{(\beta)}_L(z,\bar{z};\lambda+1).
\end{equation}
An amazing consequence of these results is that we can, following the logic used in~\cite{Loebb, Petkou, Petkou0, Petkou2}
(and in subsection~\ref{sect:2.3} to construct operator~\eqref{rl}), reduce the consideration for the arbitrary even dimension case to the two-dimensional case ($\lambda = 0$).
This can be seen as a huge advantage, since the two-dimensional integrals are known to have a very constraint form (see e.g.~\cite{Duhr:2023bku}). In particular, we focus on the results of the work~\cite{DKO}, where the notable factorization in $z$ and $\bar{z}$ was shown\footnote{Note a similar factorization for the two-dimensional two-loop master diagram~\cite{Derkachev:2022lay}.}.

\begin{proposition}
\label{pro-twodim-1}
In the case $\lambda = 0$ and  generic $\beta$, the following representations for the function $\widetilde{\Phi}^{(\beta)}_{L}(z,\bar{z}; 0)$ hold\footnote{In the work~\cite{Loebb}, representation \eqref{twodim-1} is called the Fourier-Mellin representation.}:
\begin{align}
\label{twodim-1}
&\widetilde{\Phi}_{L}^{(\beta)}(z, \bar{z}; 0) =
4^{-\beta(L + 1)}\sum_{n \in \mathbb{Z}}\int\limits_{-\infty}^{\infty} d\nu\,\dfrac{\Gamma^{L + 1}(\frac{n+1}{2} - \beta + \iu\nu)\Gamma^{L + 1}( \frac{n+1}{2} - \iu\nu)}{\Gamma^{L + 1}(\frac{n+1}{2} + \beta - \iu\nu)\Gamma^{L + 1}(\frac{n+1}{2} + \iu\nu)}z^{\iu\nu + \frac{n}{2}}\bar{z}^{\iu\nu - \frac{n}{2}}  \\
\label{twodim-2}
&=\dfrac{2\pi}{L!}4^{-\beta(L + 1)}
\partial_{\varepsilon}^L \big|_{\varepsilon = 0}
\,\varepsilon^{L + 1}\,
(z\bar{z})^{\frac{1}{2}-\varepsilon}\dfrac{\sin^{L + 1}\big(\pi(\beta + \varepsilon)\big)}{\sin^{L + 1}(\pi\varepsilon)}\,F_{L}(\beta, \varepsilon \big| z)F_{L}(\beta, \varepsilon \big| \bar{z}) \,,
\end{align}
where the function $F_{L}(\beta, \varepsilon \big| z)$
is defined by the series expansion
\begin{equation}
\label{twodim-6}
F_{L}(\beta, \varepsilon \big| z) = \sum_{k = 0}^{\infty}
\dfrac{\Gamma^{L+1}(1-\beta + k - \varepsilon)}
{\Gamma^{L+1}(1 + k - \varepsilon)}\,z^k.
    \end{equation}
\end{proposition}
{\bf Proof.} The first equality in \eqref{twodim-1}
follows from the representation \eqref{D=2f}.
To prove \eqref{twodim-2}, we use the result from Appendix~\ref{app:B}
\begin{multline*}
\sum_{n \in \mathbb{Z}}\int\limits_{-\infty}^{\infty} d\nu\,\dfrac{\Gamma^{L + 1}(\frac{n+1}{2} - \beta + \iu\nu)\Gamma^{L + 1}( \frac{n+1}{2} - \iu\nu)}{\Gamma^{L + 1}(\frac{n+1}{2} + \beta - \iu\nu)\Gamma^{L + 1}(\frac{n+1}{2} + \iu\nu)}z^{\iu\nu + \frac{n}{2}}\bar{z}^{\iu\nu - \frac{n}{2}}  =
\frac{2\pi}{L!}
\left.\partial_{\varepsilon}^{L}\right|_{\varepsilon=0}\,
(z\bar{z})^{\frac{1}{2}-\varepsilon} \\
\left(\frac{\Gamma(1-\varepsilon)\Gamma(1+\varepsilon)}
{\Gamma(\beta+\varepsilon)\Gamma(1-\beta-\varepsilon)}\right)^{L+1}\,
\sum_{p=0}^{+\infty}
\frac{\Gamma^{L+1}(1-\beta+p-\varepsilon)}
{\Gamma^{L+1}(1+p-\varepsilon)}\,z^p\,
\sum_{k=0}^{+\infty}\frac{\Gamma^{L+1}(1-\beta+k-\varepsilon)}
{\Gamma^{L+1}(1+k-\varepsilon)}\,\bar{z}^{k}.
\end{multline*}
so that expression for $\widetilde{\Phi}^{(\beta)}_L(z ,\bar{z};0)$
can be rewritten in the form
\begin{align*}
&\widetilde{\Phi}^{(\beta)}_L(z ,\bar{z}; 0)  =
4^{-\beta(L+1)}\,\frac{2\pi}{L!}
\partial_{\varepsilon}^{L}\big|_{\varepsilon = 0}
\left[\frac{\Gamma(1-\varepsilon)\Gamma(1+\varepsilon)}
{\Gamma(\beta+\varepsilon)\Gamma(1-\beta-\varepsilon)}\right]^{L+1}\,
(z\bar{z})^{\frac{1}{2}-\varepsilon}\, F_{L}(\beta\,,\varepsilon|z)\,F_{L}(\beta\,,\varepsilon|\bar{z})\,.
\end{align*}
The function $F_{L}(\beta, \varepsilon, \big| z)$ is defined in~\eqref{twodim-6}.
The factor containing $\Gamma$-functions can be simplified using the reflection property
\begin{align*}
\frac{\Gamma(1-\varepsilon)\Gamma(1+\varepsilon)}
{\Gamma(1-\beta-\varepsilon)\Gamma(\beta+\varepsilon)} =
\frac{\varepsilon \Gamma(\varepsilon)\Gamma(1-\varepsilon)}
{\Gamma(1-\beta-\varepsilon)\Gamma(\beta+\varepsilon)} = \varepsilon
\frac{\sin\pi(\beta+\varepsilon)}{\sin\pi\varepsilon},
\end{align*}
    which leads to ~\eqref{twodim-2}.
\hfill\qed

{\bf Remark 10.} Combining the result of proposition~\ref{pro1} (see eq.~\eqref{rd2}) and representation~\eqref{twodim-2}, we obtain the result for arbitrary $\lambda \in \mathbb{Z}_{>0}$ and generic $\beta$ based on the two-dimensional factorized formula
\begin{align}
\label{twodim-3}
&\widetilde{\Phi}_L^{(\beta)}(z,\bar{z}; \lambda) = \dfrac{2\pi}{L!}4^{-\beta(L + 1)}\,R_d^{\lambda}\,
\partial_{\varepsilon}^L\big|_{\varepsilon = 0}
\varepsilon^{L + 1}\,
(z\bar{z})^{\frac{1}{2}-\varepsilon}
\dfrac{\sin^{L + 1}\big(\pi(\beta + \varepsilon)\big)}{\sin^{L + 1}(\pi\varepsilon)}\,F_{L}(\beta, \varepsilon \big| z)
F_{L}(\beta, \varepsilon \big| \bar{z}).
\end{align}

{\bf Remark 11.} In the case $\beta \in \mathbb{Z}$, the factor contained sines can be simplified to exclude the dependence on~$\varepsilon$, namely,
\begin{equation}
    \dfrac{\sin\big(\pi(\beta +\varepsilon)\big)}{\sin(\pi\varepsilon)} = (-1)^{\beta}.
\end{equation}

\subsection{Consideration of conformal
ladder diagrams for $\beta=2,3,\ldots$\label{sect:3.2}}

In this subsection we consider
the case of special discrete points $\beta = 2,3,\ldots$.
The expression $\Phi^{(\beta)}_L({\sf x},{\sf y})$ \eqref{genf04b}
for the ladder diagram is a function of the complex variable $\beta$ and this
function is defined in a standard way as an analytical continuation
from the region $0 < \Re \beta < D/2$ where all integrals
in $\Phi^{(\beta)}_L({\sf x},{\sf y})$ converge absolutely.
This region is determined by two conditions.
The horizontal line on Fig.~\ref{fig2} has index
$D/2-\beta$ and the corresponding singularity of the integrand is
integrable under the condition $D/2 - \Re \beta < D/2$, which results in $\Re\beta > 0$.
The vertical line has index $\beta$ and the singularity is integrable under the condition
$\Re\beta < D/2$. Combining these conditions we obtain the region of
convergence $0 < \Re \beta < D/2$.
In this region of complex variable $\beta$ the function
$\Phi^{(\beta)}_L({\sf x},{\sf y})$ is regular. The analytical continuation from this region is a meromorphic function that has poles at the points $\beta = D/2 + k = \lambda + 1 + k$, where $k = 0,1,2,\ldots$.

Let us start with the case $D=2$ and use~\eqref{D=2a} to demonstrate the origin of singularities at a discrete set of points
$\beta = 1,2,3,\ldots$.
The contour of integration (see Remark 2) separates
two series of poles of the integrand: the first sequence at the points
$\nu_k = -\iu\frac{n+1}{2} - \iu k, k = 0, 1, \ldots$ and
the second sequence at the points
$\nu_k = -\iu \beta + \iu\frac{n+1}{2} + \iu k, k = 0, 1, \ldots$.
For real $\beta$ and $ 0 < \beta < 1$ such a contour exists for
all $n\in \mathbb{Z}_{\geq 0}$. But for $\beta \to 1$ one obtains
the pinch of the contour by two poles at $\nu = -\iu\frac{1}{2}$ and
at $\nu = \iu\frac{1}{2}-\iu\beta$ for $n=0$ which results in singularity at $\beta=1$.
For $\beta \to 2$ one obtains the pinch of the contour by two poles at $\nu = -\iu$ and at $\nu = \iu-\iu \beta$ for $n=1$ and so on.

For integer $\lambda$ the function 
$\widetilde{\Phi}_{L}^{(\beta)}(z, \bar{z}; \lambda) =
(R_d)^{\lambda}\widetilde{\Phi}_{L}^{(\beta)}(z, \bar{z}; 0)$
is obtained from the function 
$\widetilde{\Phi}_{L}^{(\beta)}(z, \bar{z}; 0)$
by applying the operator $(R_d)^{\lambda}$, which acts
on the $z$ and $\bar{z}$ variables.
It is easy to see that the operator $(R_d)^{\lambda}$ 
annihilates all functions
$(z^n+\bar{z}^n)(z\bar{z})^{{\sf i}\nu-n/2}$ for $n=1,2,\ldots,\lambda-1$ so that
for the function $\widetilde{\Phi}_{L}^{(\beta)}(z, \bar{z}; \lambda)$
the corresponding terms in the sum are absent and
the pinch of integration contour appears for $\beta = \lambda+1+k$,
where $k = 0,1,2\ldots$.
Roughly speaking, all singularities at the points $\beta = 1,2\ldots,\lambda$,
which are present in the function $\widetilde{\Phi}_{L}^{(\beta)}(z, \bar{z}; 0)$,
are annihilated by the operator $R_d^{\lambda}$ so that
the function $\widetilde{\Phi}_{L}^{(\beta)}(z, \bar{z}; \lambda)$ is regular at these points according to the regularity condition $0<\beta<\frac{D}{2} = \lambda+1$.

The expressions for the ladder integrals in dimensions $D=6,8,\ldots$
can be obtained by the application of the operator $R_d$ to
the corresponding expression for $D=4$ so
that the case $D=4$ plays the crucial role.
We have the following expression for the ladder integral in $D=4$
and arbitrary integer $\beta$:
\begin{multline}
\widetilde{\Phi}^{(\beta)}_L(z ,\bar{z};1)  =
\frac{2\pi}{L!}\,4^{-\beta(L+1)}\,
\frac{\sqrt{z \bar{z}}}{z-\bar{z}}
\left[z\partial_z-\bar{z}\partial_{\bar{z}}\right]\\
\left.\partial_{\varepsilon}^{L}\right|_{\varepsilon=0}\,
\varepsilon^{L+1}\,
\frac{\sin^{L+1}(\pi(\beta+\varepsilon))}{\sin^{L+1}(\pi\varepsilon)}\,
(z\bar{z})^{-\varepsilon}\, F_{L}(\beta\,,\varepsilon|z)\,F_{L}(\beta\,,\varepsilon|\bar{z}) = \\ \frac{2\pi}{L!}\,4^{-\beta(L+1)}\,(-1)^{\beta(L+1)}
\frac{\sqrt{z \bar{z}}}{z-\bar{z}}
\left[z\partial_z-\bar{z}\partial_{\bar{z}}\right]
\left.\partial_{\varepsilon}^{L}\right|_{\varepsilon=0}\,
\varepsilon^{L+1}\,
(z\bar{z})^{-\varepsilon}\, F_{L}(\beta\,,\varepsilon|z)\,F_{L}(\beta\,,\varepsilon|\bar{z})
\end{multline}
In what follows we present how this formula can be used in the case of $\beta = 1, 2$ and then give some remarks about the general case $\beta \in \mathbb{Z}_{>2}$.

\textbf{Case $\beta = 1$:} Let us start from the simplest example $\beta = 1$ in
order to show how everything works and to perform some cross-checks
by reproducing the results of subsection~\ref{sect:3.1}.
We have in the case $\beta = 1$ \footnote{To be rigorous, we
should note that the order of the operators $R_d$ and $\left.\partial_{\varepsilon}^{L}\right|_{\varepsilon=0}$
should be opposite. For simplicity we ignore these subtleties and hope that this will not cause any misunderstanding.}
\begin{align}
\widetilde{\Phi}^{(1)}_L(z ,\bar{z};1)  =
(-1)^{L+1}\,\frac{2\pi}{L!4^{L+1}}\,
\sqrt{z\bar{z}}\,R_d\,
\left.\partial_{\varepsilon}^{L}\right|_{\varepsilon=0}
\varepsilon^{L+1}\,
(z\bar{z})^{-\varepsilon}\, F_{L}(1\,,\varepsilon|z)\,F_{L}(1\,,\varepsilon|\bar{z}).
\end{align}
At first sight, the operation $\left.\partial_{\varepsilon}^{L}\right|_{\varepsilon= 0}
\varepsilon^{L+1}$ should produce a vanishing result, but the function
$F_{L+1}(0\,,\varepsilon|z)$ contains a singular contribution
\begin{align*}
F_{L}(1\,,\varepsilon|z) = \sum_{k=0}^{+\infty}
\frac{\Gamma^{L+1}(k-\varepsilon)}
{\Gamma^{L+1}(1+k-\varepsilon)}\,z^{k} =
\sum_{k=0}^{+\infty}
\frac{z^{k}}{(k-\varepsilon)^{L+1}} =
\frac{1}{(-\varepsilon)^{L+1}} + \sum_{k=1}^{+\infty}
\frac{z^{k}}{(k-\varepsilon)^{L+1}},
\end{align*}
so the whole expression can be transformed to
\begin{multline}
(-1)^{L+1}\,\partial_{\varepsilon}^{L}
\varepsilon^{L+1}\,
(z\bar{z})^{-\varepsilon}\, F_{L}(1\,,\varepsilon|z)\,
F_{L}(1\,,\varepsilon|\bar{z}) = \\
(-1)^{L+1}\,\partial_{\varepsilon}^{L}
\varepsilon^{L+1}\,(z\bar{z})^{\varepsilon}\,
\left[\frac{1}{(-\varepsilon)^{L+1}} + \sum_{k=1}^{+\infty}
\frac{z^{k}}{(k-\varepsilon)^{L+1}}\right]
\left[\frac{1}{(-\varepsilon)^{L+1}} + \sum_{k=1}^{+\infty}
\frac{\bar{z}^{k}}{(k-\varepsilon)^{L+1}}\right] = \\
\partial_{\varepsilon}^{L}
(z\bar{z})^{-\varepsilon}\,
\left[\frac{(-1)^{L+1}}{\varepsilon^{L+1}} +
\left(\sum_{k=1}^{+\infty}
\frac{z^{k}}{(k-\varepsilon)^{L+1}} +
\sum_{k=1}^{+\infty}
\frac{\bar{z}^{k}}{(k-\varepsilon)^{L+1}}\right) +
O(\varepsilon^{L+1})\right].
\end{multline}
At the next step, we have to apply the operator $R_d$ and after that put $\varepsilon \to 0$.

It should be noted that this expression essentially coincides
with the expression for the ladder diagram in $D=2$ and, therefore, has a singularity at $\beta=1$. This singularity manifests itself in the previous formula -- 
it is the contribution $\frac{(-1)^{L+1}}{\varepsilon^{L+1}}$ which 
is singular as $\varepsilon \to 0$. The convergence condition $0 < \beta < D/2 =1$ is violated in this case so that
the origin of the singularity is clear.
The ladder diagram in $D=4$ is obtained by applying the operator
$R_d$ to the expression for the ladder diagram in $D=2$.
In $D=4$, the convergence condition $0 < \beta < D/2 = 2$ is satisfied, so the operator $R_d$ must annihilate all singularities, and as we can see this is indeed the case. 
The last term $O(\varepsilon^{L+1})$ is annihilated by the operator
$\left.\partial_{\varepsilon}^{L}\right|_{\varepsilon=0}$ so that one obtains
\begin{align}
\widetilde{\Phi}^{(1)}_L(z ,\bar{z};1)  =
\frac{2\pi}{L!4^{L+1}}\,
\sqrt{z\bar{z}}\,R_d\,
\left.\partial_{\varepsilon}^{L}\right|_{\varepsilon=0}
(z\bar{z})^{-\varepsilon}\,\left(\sum_{k=1}^{+\infty}
\frac{z^{k}}{(k-\varepsilon)^{L+1}} +
\sum_{k=1}^{+\infty}
\frac{\bar{z}^{k}}{(k-\varepsilon)^{L+1}}\right)
\end{align}
After calculation of the needed derivative
\begin{align*}
&\left.\partial_{\varepsilon}^{L}\right|_{\varepsilon=0}
(z\bar{z})^{-\varepsilon}\,
\sum_{k=1}^{+\infty}
\frac{z^{k}}{(k-\varepsilon)^{L+1}} =
\sum_{p=0}^{L}\binom{L}{p}\,
\left.\partial_{\varepsilon}^{p}\,
(z\bar{z})^{-\varepsilon}\,
\partial_{\varepsilon}^{L-p}\,\sum_{k=1}^{+\infty}
\frac{z^{k}}{(k-\varepsilon)^{L+1}}\right|_{\varepsilon=0} = \\
&\sum_{p=0}^{L}\binom{L}{p}\,(-1)^{p}\,\log^{p}(z\bar{z})\,
\frac{(2L-p)!}{L!}\,\Li_{2L+1-p}(z)  =
\sum_{p=0}^{L}\frac{(-1)^{p}(2L-p)!}{p!(L-p)!}\,
\log^{p}(z\bar{z})\,\Li_{2L+1-p}(z),
\end{align*}
we reproduce \eqref{der-D4}
\begin{align*}
\widetilde{\Phi}^{(1)}_L(z ,\bar{z};1)  =&
\frac{2\pi}{L!4^{L+1}}\,
\sqrt{z \bar{z}}\,R_d\,
\sum_{p=0}^{L}\frac{(-1)^{p}(2L-p)!}{p!(L-p)!}\,
\log^{p}(z\bar{z})\,\left[\Li_{2L+1-p}(z)+\Li_{2L+1-p}(\bar{z})\right] \notag\\ =&
\frac{2\pi}{L!4^{L+1}}\,
\frac{\sqrt{z \bar{z}}}{z-\bar{z}}\,
\sum_{p=0}^{L}\frac{(-1)^{p}(2L-p)!}{p!(L-p)!}\,
\log^{p}(z\bar{z})\,\left[\Li_{2L-p}(z)-\Li_{2L-p}(\bar{z})\right].
\end{align*}
\textbf{Case $\beta = 2$:} Now we proceed to the next example $\beta=2$.
The convergence condition $0 < \beta < D/2$ is valid starting from $D=6$
so that we put $\lambda = 2$ and get from~\eqref{twodim-3}
\begin{align}
\widetilde{\Phi}^{(2)}_L(z ,\bar{z};2)  =
\frac{2\pi}{L!}\,4^{-2(L+1)}\,
\sqrt{z \bar{z}}\,R_d^2\,
\left.\partial_{\varepsilon}^{L}\right|_{\varepsilon=0}
\varepsilon^{L+1}\,
(z\bar{z})^{-\varepsilon}\, F_{L}(2\,,\varepsilon|z)\,
F_{L}(2\,,\varepsilon|\bar{z}),
\end{align}
where
\begin{align}
F_{L}(2\,,\varepsilon|z) =& \sum_{k=0}^{+\infty}
\frac{\Gamma^{L+1}(k-1-\varepsilon)}
{\Gamma^{L+1}(k+1-\varepsilon)}\,z^{k} =\,
\sum_{k=0}^{+\infty}
\frac{z^{k}}{(k-1-\varepsilon)^{L+1}\,(k-\varepsilon)^{L+1}} \notag \\ =&
\frac{\phi(z)}{\varepsilon^{L+1}} +
\sum_{k=2}^{+\infty}
\frac{z^{k}}{(k-1-\varepsilon)^{L+1}\,(k-\varepsilon)^{L+1}},
\end{align}
and
\begin{align*}
\phi(z) = \frac{1}{(1+\varepsilon)^{L+1}} +
\frac{z}{(\varepsilon-1)^{L+1}}.
\end{align*}
All calculations are similar to the case $\beta=1$, namely
\begin{multline*}
\partial_{\varepsilon}^{L}
\varepsilon^{L+1}\,
(z\bar{z})^{-\varepsilon}\, F_{L}(2\,,\varepsilon|z)\,
F_{L}(2\,,\varepsilon|\bar{z}) =
\partial_{\varepsilon}^{L}
\varepsilon^{L+1}\,(z\bar{z})^{-\varepsilon}\,\\
\left[
\frac{\phi(z)}{\varepsilon^{L+1}} +
\sum_{k=2}^{+\infty}
\frac{z^{k}}{(k-1-\varepsilon)^{L+1}\,(k-\varepsilon)^{L+1}}
\right]
\left[
\frac{\phi(\bar{z})}{\varepsilon^{L+1}} +
\sum_{k=2}^{+\infty}
\frac{\bar{z}^{k}}{(k-1-\varepsilon)^{L+1}\,(k-\varepsilon)^{L+1}}
\right] \to \\
\partial_{\varepsilon}^{L}\,
(z\bar{z})^{-\varepsilon}\,
\left[
\phi(\bar{z})\,\sum_{k=2}^{+\infty}\frac{z^{k}}{(k-1-\varepsilon)^{L+1}
\,(k-\varepsilon)^{L+1}} +
\phi(z)\,\sum_{k=2}^{+\infty}\frac{\bar{z}^{k}}
{(k-1-\varepsilon)^{L+1}\,(k-\varepsilon)^{L+1}}\right],
\end{multline*}
where in the last line we have removed two contributions.
First of all, it is the singular contribution of the form $\frac{1}{\varepsilon^{L+1}}\phi(z)\phi(\bar{z})$ which does
not contribute to the final answer because
it is annihilated by the operator $R_d^2$.
The second contribution is $O(\varepsilon^{L+1})$ and
it is annihilated by the operator $\left.\partial_{\varepsilon}^{L}\right|_{\varepsilon=0}$.
Collecting all factors together we obtain the following expression:
\begin{multline}
\widetilde{\Phi}^{(2)}_L(z ,\bar{z};2)  =
\frac{2\pi}{L!}\,4^{-2(L+1)}\,
\sqrt{z \bar{z}}\,R_d^2\,
\left.\partial_{\varepsilon}^{L}\right|_{\varepsilon=0}\,
(z\bar{z})^{-\varepsilon}\,\\
\left[
\phi(\bar{z})\,\sum_{k=2}^{+\infty}\frac{z^{k}}{(k-1-\varepsilon)^{L+1}
\,(k-\varepsilon)^{L+1}} +
\phi(z)\,\sum_{k=2}^{+\infty}\frac{\bar{z}^{k}}{(k-1-\varepsilon)^{L+1}\,
(k-\varepsilon)^{L+1}}\right].
\end{multline}
\textbf{Case $\beta \in \mathbb{Z}_{>2}$: } The generalization of the above calculations is rather straightforward.
For $\beta\in \mathbb{Z}_{>2}$ the convergence condition
$0 < \beta < D/2$ is fulfilled starting from $D=2\beta+2$ so that
$\lambda  = \beta$ and
one obtains the following representation for the~\eqref{twodim-3}:
\begin{multline}
\widetilde{\Phi}^{(\beta)}_L(z ,\bar{z};\beta)  =
(-1)^{\beta(L+1)}\,\frac{2\pi}{L!}\,4^{-\beta(L+1)}\,
\sqrt{z \bar{z}}\,R_d^{\beta}\,
\left.\partial_{\varepsilon}^{L}\right|_{\varepsilon=0}\,
(z\bar{z})^{-\varepsilon}\,\\
\left[
\phi_{\beta}(\bar{z})\,\sum_{k=\beta}^{+\infty}\frac{z^{k}}
{\prod_{p=0}^{\beta-1}(k-p-\varepsilon)^{L+1}} +
\phi_{\beta}(z)\,\sum_{k=\beta}^{+\infty}
\frac{\bar{z}^{k}}{\prod_{p=0}^{\beta-1}(k-p-\varepsilon)^{L+1}}\right],
\end{multline}
where we omitted all the contributions annihilated by the operator $R_d^{\beta}$ and $\left.\partial_{\varepsilon}^{L}\right|_{\varepsilon=0}$, and the polynomial $\phi_{\beta}(z)$ is defined by the formula
\begin{align}
\phi_{\beta}(z) = (-1)^{L+1}\,\sum_{k=0}^{\beta-1} z^{k}\,
\prod\limits_{\substack{p=0\\p\neq k}}^{\beta-1}(k-p-\varepsilon)^{-L-1}.
\end{align}

\vspace{0.5cm}

\section{Conclusions}

In this work, we have studied the family of conformal four-point ladder diagrams. This series of integrals was first calculated in the four-dimensional case with physical propagator powers~\cite{Usyukina:1992jd, Usyukina:1993ch} and continued to attract attention due to the rich underlying structure. The generalized class of four-dimensional diagrams was later shown to have remarkable analytic properties~\cite{Drum}. The generalization to higher even dimensions turned out to be suitable for application of the theory of graphical functions~\cite{Borinsky:2021gkd} with the explicit result obtained in~\cite{Borinsky:2022lds}. Besides their interest as computational challenges, these integrals were found to be connected with twisted partition functions~\cite{Petkou0, Petkou, Petkou2}. Significant progress was also made in understanding the recursive relations connecting diagrams in different dimensions and with different loop numbers~\cite{Loebb}.
\vspace{0.2cm}

Using the iterative structure of the ladder diagrams,  we showed in~\cite{DIS2} that the graph-building operator method together with conformal quantum mechanics provides an explicit answer for the integrals in arbitrary dimension. 
The present work can be seen as a continuation of those studies. In the paper, we have explicitly derived that the representation obtained in~\cite{DIS2} satisfies dimensional and loop shift identities for arbitrary dimensions and general propagator powers described by the parameter $\beta$. Additionally, in two-dimensions and generic $\beta$ we observed a remarkable factorization analogous to the two-dimensional fishnet diagrams~\cite{DKO}, which by the dimensional shift identity can be translated to all even dimensions. Specializing further, we showed that for $\beta = 1$ (which in the four-dimensional case corresponds to the physical propagator powers) and even $D$ our representation can be rewritten in terms of classical polylogarithms with the rational functions. For higher integer values of $\beta$ the representation based on two-dimensional factorization turned out to be more suitable and was studied. Notably, it naturally provides a regularization for the infrared singularities that arise when $\beta = \frac{D}{2} + k$ with $k \in \mathbb{Z}_{\ge 0}$.
\vspace{0.2cm}

Our results establish clear links between the operator-based construction of~\cite{DIS2} and alternative approaches to conformal four-point integrals. This not only enriches the graph-building operator method but also offers potential benefits for alternative methods, particularly given by the well-studied properties of the  operator $H_{\alpha}$, reflecting the underlying conformal and integrable structures. We hope that these insights may contribute to revealing further internal symmetries such as, e.g., antipodal self-duality~\cite{Dixon:2025zwj}, and studying of more general families of conformal integrals~\cite{He:2025lzd}.

\vspace{0.5cm}

\appendix

\section*{Acknowledgments}

We are grateful to A.C.~Petkou for drawing our attention to the problem of evaluating conformal
ladder diagrams in diverse dimensions. We would also like to thank A.I.~Davydychev, A.V.~Kotikov and A.C.~Petkou  for useful comments and stimulating discussions.


\appendix

\section{The limit $\lambda \to 0$ for the function
$\widetilde{\Phi}^{(\beta)}_L(z ,\bar{z};\lambda)$\label{app:C}}

To get rid of the normalizing factor in (\ref{D=2})
during calculations, we will perform the limit $\lambda \to 0$
for the renormalized function (\ref{beta0}), (\ref{D=D}).

The generating function of the Chebyshev polynomials of the first kind
\be
\lb{cheb}
T_n (\cos \theta) = \cos n \theta \; ,
\ee
 is written as (see \cite{BatErd2}, section 10.11)
\be
\lb{cheb01}
\ln(1-2tr+t^2) = -2\sum_{n=1}^\infty T_n(r) \frac{t^n}{n} \; .
\ee
Thus, taking into account the generating function (\ref{derk}) of
the Gegenbauer polynomials $C^{(\lambda)}_n(r)$, we find the
expansion of $C^{(\lambda)}_n(r)$ as $\lambda \to 0$:
 \be
 \lb{cheb02}
 C^{(\lambda)}_0(r) = 1 \; ,
 \;\;\;\;\;
 C^{(\lambda)}_n(r) = \lambda \frac{2}{n} T_n(r) + \lambda^2 ... \;
 \;\;\;\;\; (\forall n \geq 1) \; .
 \ee
 We use this expansion in \eqref{D=D} and derive
 \begin{equation}
\label{twodim-4b}
\begin{array}{l}
   \left. \widetilde{\Phi}^{(\beta)}_L(z ,\bar{z};\lambda)
   \right|_{\lambda \to 0} = \\ [0.2cm]
  =  \dfrac{\Gamma(\lambda)}{(z \bar{z})^{\lambda/2}} \left(
    \lambda
    \, \int\limits_{-\infty}^{+\infty} d\nu
     \dfrac{(z\bar{z})^{{\sf i}\nu}}{\left[\tau_{0,\nu}(\beta;0)\right]^{L+1}}
      + 2\, \lambda \, \sum\limits_{n=1}^\infty
    T_n\left(r\right)
    \int\limits_{-\infty}^{+\infty} d\nu
     \dfrac{(z\bar{z})^{{\sf i}\nu}}{
 \left[\tau_{n,\nu}(\beta;0)\right]^{L+1}}
 + \lambda^2 ... \right) ,
     \end{array}
     \ee
     where, according to (\ref{cheb})
     taking $z = |z|e^{{\sf i}\theta}$, we have
 $$
 \begin{array}{c}
 r = \dfrac{z+\bar{z}}{2\sqrt{z\bar{z}}}=
 \dfrac{1}{2}\Bigl(
  \left(z/\bar{z}\right)^{1 \over 2} +
  \left(\bar{z}/z\right)^{1 \over 2}
 \Bigr) = \cos \theta
 \; , \;\;\;\; \\ [0.2cm]
 T_n(r) = \cos n \theta  =
 \dfrac{1}{2}\Bigl( \left(z/\bar{z}\right)^{n \over 2} +
  \left(\bar{z}/z\right)^{n \over 2}\Bigr) =
   \dfrac{z^n + \bar{z}^n}{2\, (z\bar{z})^{n/2}} \; .
 \end{array}
 $$
   Therefore we obtain (cf. (\ref{D=2}), (\ref{D=2a}))
     \be
     \label{twodim-4d}
     \begin{array}{c}
    \widetilde{\Phi}^{(\beta)}_L(z ,\bar{z};0)
     = \int\limits_{-\infty}^{+\infty} d\nu
     \dfrac{(z\bar{z})^{{\sf i}\nu}}{\left[\tau_{0,\nu}(\beta;0)\right]^{L+1}}
     + 2\, \sum\limits_{n=1}^\infty
    T_n\left(r\right)
    \int\limits_{-\infty}^{+\infty} d\nu
     \dfrac{(z\bar{z})^{{\sf i}\nu}}{
     \left[\tau_{n,\nu}(\beta;0)\right]^{L+1}} = \\ [0.5cm]
  = \int\limits_{-\infty}^{+\infty} d\nu
     \dfrac{(z\bar{z})^{{\sf i}\nu}}{\left[\tau_{0,\nu}(\beta;0)\right]^{L+1}}
     + \sum\limits_{n=1}^\infty
    \left(z^n + \bar{z}^n\right)
    \int\limits_{-\infty}^{+\infty} d\nu
     \dfrac{(z\bar{z})^{{\sf i}\nu-n/2}}{
     \left[\tau_{n,\nu}(\beta;0)\right]^{L+1}}    \; .
      \end{array}
    \ee
 Note that the first term in the r.h.s. of \eqref{twodim-4d}
 is the  zero mode of the operator $R_d$
 defined in \eqref{rd}.

It is noteworthy that expression (\ref{twodim-4d})
can be written in a more compact form by means of
the symmetry property of the eigenvalue (\ref{tau1})
for $\lambda =0$:
\begin{equation}
\label{symtau}
    \tau_{n,\nu}(\beta;0) = \tau_{-n,\nu}(\beta;0) \; .
\end{equation}
Then, we perform the change of variables $n \mapsto - n $ in the
right-hand side of (\ref{twodim-4d}) in the
sum containing $\bar{z}^{n}$, use property
(\ref{symtau}), and finally obtain the concise formula
\begin{equation}
\lb{D=2b}
\widetilde{\Phi}^{(\beta)}_L(z ,\bar{z};0)    =
 \sum_{n \in \mathbb{Z}}
\int\limits_{-\infty}^{+\infty} d\nu
\dfrac{z^{{\sf i}\nu+\frac{n}{2}}\,\bar{z}^{{\sf i}\nu-\frac{n}{2}}}
 {\left[\tau_{n,\nu}(\beta;0)\right]^{L+1}}\,.
\end{equation}
The symmetry property (\ref{symtau}) for the expression
\begin{equation}
\lb{tau05}
\tau_{n,\nu}(\beta;0) =
4^\beta \frac{\Gamma(\frac{n+1}{2}+\beta -{\sf i}\nu)}{
\Gamma(\frac{n+1}{2}-\beta + {\sf i}\nu)} \;
\frac{\Gamma(\frac{n+1}{2}+{\sf i}\nu)}{
\Gamma(\frac{n+1}{2}-{\sf i}\nu)}
\end{equation}
can be easily checked.
Indeed, using the reflection formula
$\Gamma(x)\Gamma(1-x) = \pi /\sin(\pi x)$ for the
$\Gamma$-function, one can show that
\begin{align*}
\frac{\Gamma(\frac{n+1}{2}+\beta -{\sf i}\nu)}{
\Gamma(\frac{n+1}{2}-\beta + {\sf i}\nu)} =
(-1)^{n}
\frac{\Gamma(\frac{1-n}{2}+\beta-{\sf i}\nu)}{
\Gamma(\frac{1-n}{2}-\beta+ {\sf i}\nu)}
\;\;\; \stackrel{\beta=0}{\Rightarrow}  \;\;\;
\frac{\Gamma(\frac{n+1}{2} -{\sf i}\nu)}{
\Gamma(\frac{n+1}{2} + {\sf i}\nu)} =
(-1)^{n}
\frac{\Gamma(\frac{1-n}{2}-{\sf i}\nu)}{
\Gamma(\frac{1-n}{2}+ {\sf i}\nu)},
\end{align*}
from which the property of symmetry
(\ref{symtau}) immediately follows.

\section{Explicit check of the loop shift identity for $\lambda = 1, \beta = 1$\label{app:d}}

This appendix is devoted to the explicit check of identity~\eqref{C14}. We use the representation~\eqref{der-D4} for the function $\tilde{\Phi}^{(1)}_{L}(z, \bar{z}; 1)$ in terms of classical polylogarithms
\begin{align*}
\widetilde{\Phi}^{(1)}_L(z ,\bar{z};1) =
\frac{2\pi}{L!4^{L+1}}\,
\frac{\sqrt{z \bar{z}}}{z-\bar{z}}\,
\sum_{p=0}^{L}\frac{(-1)^{p}(2L-p)!}{p!(L-p)!}\,
\log^{p}(z\bar{z})\,\left[\Li_{2L-p}(z)-\Li_{2L-p}(\bar{z})\right].
\end{align*}
Applying the operator $\left(R^{(1)}_{\ell}(1)\right)^{-1}$ in the form~\eqref{RBL} with $\beta = 1$ we have
\begin{align*}
-4\,\frac{\sqrt{z \bar{z}}}{z-\bar{z}}&\,
z\partial_z\,\bar{z}\partial_{\bar{z}}\,
\frac{z-\bar{z}}{\sqrt{z\bar{z}}}\,\widetilde{\Phi}^{(1)}_L(z ,\bar{z};1) \notag \\
=&-\frac{2\pi}{L!4^{L}}\,\frac{\sqrt{z\bar{z}}}{z-\bar{z}}\,
z\partial_z\,\bar{z}\partial_{\bar{z}}\,\sum_{p=0}^{L}\frac{(-1)^{p}(2L-p)!}{p!(L-p)!}\,
\log^{p}(z\bar{z})\,\left[\Li_{2L-p}(z)-\Li_{2L-p}(\bar{z})\right] \notag\\
=&\frac{2\pi}{(L-1)!4^{L}}\,\frac{\sqrt{z\bar{z}}}{z-\bar{z}}\,
 \sum_{p=0}^{L-1}\frac{(-1)^{p}(2L-p-2)!}{p!(L-p-1)!}\,
\log^{p}(z\bar{z})\,\left[\Li_{2L-p-2}(z)-\Li_{2L-p-2}(\bar{z})\right]\notag \\ =& \widetilde{\Phi}^{(1)}_{L-1}(z ,\bar{z};1),
\end{align*}
where we used the following formula:
\begin{align*}
z\partial_z\,\bar{z}\partial_{\bar{z}}\,
\log^{p}(z\bar{z})\,\Li_{2L-p}(z) =
p(p-1)\log^{p-2}(z\bar{z})\,\Li_{2L-p}(z) +
p\log^{p-1}(z\bar{z})\,\Li_{2L-p-1}(z)
\end{align*}
so that
\allowdisplaybreaks{
\begin{multline*}
z\partial_z\,\bar{z}\partial_{\bar{z}}\,
\sum_{p=0}^{L}\frac{(-1)^{p}(2L-p)!}{p!(L-p)!}\,
\log^{p}(z\bar{z})\,\Li_{2L-p}(z) = \\
=\sum_{p=0}^{L-2}\frac{(-1)^{p}(2L-p-2)!}{p!(L-p-2)!}\,
\log^{p}(z\bar{z})\,\Li_{2L-p-2}(z)   -
\sum_{p=0}^{L-1}\frac{(-1)^{p}(2L-p-1)!}
{p!(L-p-1)!}\,\log^{p}(z\bar{z})\,\Li_{2L-p-2}(z) \\
=\sum_{p=0}^{L-1}\frac{(-1)^{p}(2L-p-2)!}{p!(L-p-1)!}\,
\left(L-p-1 -(2L-p-1) \right)\,
 \log^{p}(z\bar{z})\,\Li_{2L-p-2}(z)  \\=
(-L)\sum_{p=0}^{L-1}\frac{(-1)^{p}(2L-p-2)!}{p!(L-p-1)!}\,
 \log^{p}(z\bar{z})\,\Li_{2L-p-2}(z).
\end{multline*}
}

\section{Symmetrical polynomials in $z, \bar{z}$ in the case $\beta = 1$\label{app:a}}
In this appendix, we discuss the properties of the rational functions arising as coefficients of classical polylogarithms in the answer for ladder diagrams with arbitrary positive integer $\lambda$ and $\beta = 1$ (this corresponds to the ladder diagrams~Fig.~\ref{fig1} in $D = 2\lambda + 2$ dimensions with indices $1$ on the horizontal lines and $D/2 - 1$ on the vertical lines). In what follows we summarize and generalize the properties noted in the examples in section~\ref{sect:3.1}.
\begin{proposition}\lb{pro3}
  In the case $\lambda \in \mathbb{Z}_{>0}$, the function $\Sigma_s^{(\lambda)}(z, \bar{z})$ defined in~\eqref{der02} can be expressed in the form
  \begin{equation}
      \label{shu04}
      \Sigma_s^{(\lambda)}(z, \bar{z}) = \sum_{k = 0}^{\lambda - 1}\dfrac{G^{(\lambda)}_{k}(z, \bar{z})}{(z - \bar{z})^{2\lambda - 1 - k}}\Li_{s - k}(z) + (z \leftrightarrow \bar{z}),
  \end{equation}
  where the polynomial $G_k^{(\lambda)}(z, \bar{z})$ can be written in the form
  \begin{equation}
  \label{shu05}
      G_k^{(\lambda)}(z, \bar{z}) = \dfrac{(z - \bar{z})^{2\lambda - 1 - k}}{(\lambda - 1)!}\dfrac{1}{k!}\partial_t^{\lambda - 1}\dfrac{(1 + t)^{\lambda - 1}\log^k\left(1 + t\right)}{((1 + t)z - \bar{z})^\lambda}\Bigg|_{t = 0}.
  \end{equation}
\end{proposition}
{\bf Proof.}
We start the relation~\eqref{der16} and use the representation in the r.h.s. of~\eqref{der10} to express the operator $P_{\lambda}(z\partial_z)$ (below we use the notation from~\eqref{rd02})
\begin{equation}
\label{app01}
    \Sigma_s^{\prime (\lambda)}(z, \bar{z}) = \dfrac{1}{\Gamma(\lambda)}\partial_z^{\lambda - 1}\dfrac{z^{\lambda - 1}\Li_s(z)}{(z - \bar{z})^\lambda} = \dfrac{1}{\Gamma(\lambda)}\sum_{n = 0}^{\lambda - 1}\binom{\lambda - 1}{n}\partial_z^{\lambda - 1 - n}\dfrac{z^{\lambda - 1}}{(z - \bar{z})^{\lambda}}\partial_z^n\Li_s(z).
\end{equation}
In order to use the property~\eqref{shu02} to express the derivative of polylog, we rewrite the derivative in the form
\begin{equation*}
    \partial_z^n = \dfrac{1}{z^n}\left(z\partial_z - n + 1\right)\ldots(z\partial_z - 1)z\partial_z,
\end{equation*}
which follows from the fact $z^{-k}(z\partial_z - k + 1) = \partial_zz^{-k + 1}$. Introducing an auxiliary variable $\alpha$ we rewrite
\begin{equation}
    \partial_z^{n} = \dfrac{1}{z^n}\partial^n_\alpha\alpha^{z\partial_z}\Big|_{\alpha = 1}.
\end{equation}
Putting this relation in~\eqref{app01} and expanding $\alpha^{z\partial_z}$ in Taylor series, we get
\begin{equation}
    \Sigma_s^{\prime(\lambda)}(z, \bar{z}) = \dfrac{1}{\Gamma(\lambda)}\sum_{k = 0}^{\infty}\dfrac{(z\partial_z)^k\Li_s(z)}{k!}\sum_{n = 0}^{\lambda - 1}\binom{\lambda - 1}{n}\partial_z^{\lambda - 1 - n}\dfrac{z^{\lambda - 1}}{(z - \bar{z})^{\lambda}}\dfrac{\partial^n_\alpha}{z^n}\log^k\alpha\Big|_{\alpha = 1}.
\end{equation}
Now we use~\eqref{shu02} and calculate the sum over $n$ which results in
\begin{equation}
\label{app02}
    \Sigma_s^{\prime(\lambda)}(z, \bar{z}) = \dfrac{1}{\Gamma(\lambda)}\sum_{k = 0}^{\infty}\dfrac{\Li_{s- k}(z)}{k!}\left(\partial_z + \dfrac{1}{x}\partial_\alpha\right)^{\lambda - 1}\dfrac{z^{\lambda - 1}}{(z- \bar{z})^\lambda}\log^k\alpha\Big|_{\alpha = 1, x = z}.
\end{equation}
Note that we introduce an auxiliary variable $x$ to underline that one should treat operators in the brackets as commuting and put $x = z$ after applying all derivatives. Also note that if $p > \lambda - 1$, the corresponding terms in the sum are nullified after substituting $\alpha = 1$. Expression~\eqref{app02} allows us to single out the contribution proportional to the polylogs and get the following representation for the polynomial~\eqref{shu05}:
\begin{equation*}
    G_k^{(\lambda)}(z, \bar{z}) =  \dfrac{(z - \bar{z})^{2\lambda - 1 - k}}{(\lambda - 1)!}\dfrac{1}{k!}\left(\partial_z + \dfrac{1}{x}\partial_\alpha\right)^{\lambda - 1}\dfrac{z^{\lambda - 1}}{(z - \bar{z})}\log^k\alpha\Big|_{\alpha = 1, x = z}.
\end{equation*}
This formula can be simplified by introducing another auxiliary variable $t$
\begin{align*}
     G_k^{(\lambda)}(z, \bar{z}) =& \dfrac{(z - \bar{z})^{2\lambda - 1 - k}}{(\lambda - 1)!}\dfrac{1}{k!}\partial_t^{\lambda - 1}e^{t\big(\partial_z + \frac{1}{x}\partial_\alpha\big)}\dfrac{z^{\lambda - 1}}{(z - \bar{z})}\log^k\alpha\Big|_{\alpha = 1, x = z, t = 0} \notag \\
     =& \dfrac{(z - \bar{z})^{2\lambda - 1 - k}}{(\lambda - 1)!}\dfrac{1}{k!}\partial_t^{\lambda - 1}\dfrac{(z + t)^{\lambda - 1}}{(z - \bar{z} + t)^{\lambda}}\log^k\left(\alpha + \dfrac{t}{x}\right)\Big|_{\alpha = 1, x = z, t = 0} \notag \\
     =& \dfrac{(z - \bar{z})^{2\lambda - 1 - k}}{(\lambda - 1)!}\dfrac{1}{k!}\partial_t^{\lambda - 1}\dfrac{(z + t)^{\lambda - 1}}{(z - \bar{z} + t)^\lambda}\log^k\left(1 + \dfrac{t}{z}\right)\Big|_{t = 0}.
\end{align*}
Rescaling the auxiliary variable $t \mapsto tz$, we conclude~\eqref{shu05}.
\hfill\qed
\vspace{0.3cm}
\begin{proposition}\lb{pro4}
    THe function $G_k^{(\lambda)}(z, \bar{z})$ defined in~\eqref{shu05} is a homogeneous polynomial of degree $\lambda - 1 - k$ symmetrical under the transformation $z \leftrightarrow \bar{z}$ if $\lambda \in \mathbb{Z}_{>0}$ and $k = 0, \ldots, \lambda - 1$.
\end{proposition}
\textbf{Proof.} We organize the proof of this proposition in three steps. First, we show that $G_k^{(\lambda)}(z, \bar{z})$ is a polynomial. Indeed, application of the derivative with respect to $t$ can be rewritten as
\begin{equation*}
    G_k^{(\lambda)}(z, \bar{z}) = \dfrac{(z - \bar{z})^{2\lambda - 1 - k}}{(\lambda - 1)!}\dfrac{1}{k!}\sum_{n = 0}^{\lambda - 1}\binom{\lambda - 1}{n}\partial_t^n\Big[(1 + t)^{\lambda - 1}\log^k(1 + t)\Big]\Big|_{t = 0}\partial_t^{\lambda - 1 - n}\dfrac{1}{((1 + t)z - \bar{z})^\lambda}\Big|_{t = 0}.
\end{equation*}
Note that the first term
\begin{equation*}
    \partial_t^n\Big[(1 + t)^{\lambda - 1}\log^k(1 + t)\Big]\Big|_{t = 0} = 0, \qquad \text{if}~~n \le k.
\end{equation*}
Thus, the highest power of $(z - \bar{z})$ in the denominator is $\lambda - 1- k + \lambda = 2\lambda - 1 - k$, which is precisely canceled by the prefactor.
Second, we show that the polynomial $G_k^{(\lambda)}(z, \bar{z})$ is homogeneous. Introducing an arbitrary parameter $\mu \in \mathbb{R}$, we conclude
\begin{equation*}
    G_k^{(\lambda)}(\mu z, \mu\bar{z}) = \dfrac{(\mu z - \mu\bar{z})^{2\lambda - 1 - k}}{(\lambda - 1)!}\dfrac{1}{k!}\partial_t^{\lambda - 1}\dfrac{(1 + t)^{\lambda - 1}\log^k\left(1 + t\right)}{((1 + t)\mu z - \mu\bar{z})^\lambda}\Bigg|_{t = 0} = \mu^{\lambda - 1 -k}G_k^{(\lambda)}(z, \bar{z}),
\end{equation*}
so the polynomial $G_k^{(\lambda)}(z, \bar{z})$ is indeed homogeneous with the degree $\lambda - 1 - k$.

As the last step we show that $G_k^{(\lambda)}(z, \bar{z}) = G_k^{(\lambda)}(\bar{z}, z)$. In order to address the derivative at $t = 0$, we rewrite~\eqref{shu05} as a Cauchy integral
\begin{equation}
\label{app04}
    G_k^{(\lambda)}(z, \bar{z}) = (z - \bar{z})^{2\lambda - k - 1}\dfrac{1}{2\pi {\sf i} k!}\oint\limits_{\gamma}\dfrac{dt}{t^\lambda}\dfrac{(1 + t)^{\lambda - 1}\log^k(1 + t)}{\big((1 + t)z - \bar{z}\big)^\lambda},
\end{equation}
where $\gamma$ is the infinitesimally small contour around $t = 0$. Now we change the variables
\begin{align}
\label{app03}
    t \mapsto -\dfrac{t}{1 + t},
\end{align}
which leads to
\begin{align*}
    1 + t \mapsto \frac{1}{1 + t}, && dt \mapsto -\dfrac{dt}{(1 + t)^2}, && \operatorname{Log}(1 + t) \mapsto -\operatorname{Log}(1 + t).
\end{align*}
Also note that the contour of integration $\gamma$ maps onto itself. Applying the change of variables~\eqref{app03} to ~\eqref{app04}, we get
\begin{align*}
    G_k^{(\lambda)}(z, \bar{z}) =& - (z - \bar{z})^{2\lambda - k - 1}\dfrac{1}{2\pi{\sf i} k!}\oint\limits_{\gamma}\dfrac{dt}{(1 + t)^2}\dfrac{(1 + t)^\lambda}{t^\lambda}\dfrac{(-1)^{k + \lambda}}{(1 + t)^{\lambda - 1}}\dfrac{\log^k(1 + t)}{\big(\frac{z}{1 + t} - \bar{z}\big)^\lambda} \notag \\
    =& (-1)^{k - 1} (z - \bar{z})^{2\lambda - k - 1}\dfrac{1}{2\pi {\sf i} k!}\oint\limits_{\gamma}(-1)^\lambda\dfrac{dt}{t^\lambda}\dfrac{(1 + t)^{\lambda - 1}\log^k(1 + t)}{\big(z - (1 + t)\bar{z}\big)^\lambda}.
\end{align*}
Changing the order of terms in the brackets, we arrive at
\begin{equation*}
    G_k^{(\lambda)}(z, \bar{z}) =  (\bar{z} - z)^{2\lambda - k - 1}\dfrac{1}{2\pi {\sf i} k!}\oint\limits_{\gamma}\dfrac{dt}{t^\lambda}\dfrac{(1 + t)^{\lambda - 1}\log^k(1 + t)}{\big((1 + t)\bar{z} - z\big)^\lambda} = G_k^{(\lambda)}(\bar{z}, z),
\end{equation*}
where we used that $\lambda \in \mathbb{Z}_{>0}$, so $(-1)^{2\lambda} = 1$.
\hfill\qed

\section{Derivation of factorization in two-dimensional case\label{app:B}}

This appendix is devoted to the calculation of integral over $\nu$ in~\eqref{twodim-1} and further factorization in $z$ and $\bar{z}$. In principal, one can find a detailed derivation in~\cite[Section 5]{DKO}, but for convenience we repeat it here.
The first step is the calculation of the integral over $\nu$ by evaluating residues,
which results in the following expression
\begin{multline}
\sum_{n \in \mathbb{Z}}\int\limits_{-\infty}^{\infty} d\nu\,\dfrac{\Gamma^{L + 1}(\frac{n+1}{2} - \beta + \iu\nu)\Gamma^{L + 1}( \frac{n+1}{2} - \iu\nu)}{\Gamma^{L + 1}(\frac{n+1}{2} + \beta - \iu\nu)\Gamma^{L + 1}(\frac{n+1}{2} + \iu\nu)}z^{\iu\nu + \frac{n}{2}}\bar{z}^{\iu\nu - \frac{n}{2}}  =
\frac{2\pi}{L!}
\left.\partial_{\varepsilon}^{L}\right|_{\varepsilon=0}\,
(z\bar{z})^{\frac{1}{2}-\varepsilon} \\
\left(\frac{\Gamma(1-\varepsilon)\Gamma(1+\varepsilon)}
{\Gamma(\beta+\varepsilon)\Gamma(1-\beta-\varepsilon)}\right)^{L+1}\,
\sum_{n\in \mathbb{Z}}\sum_{k=0}^{+\infty}
\frac{\Gamma^{L+1}(1-\beta+n+k-\varepsilon)}
{\Gamma^{L+1}(1+n+k-\varepsilon)}
\frac{\Gamma^{L+1}(1-\beta+k-\varepsilon)}
{\Gamma^{L+1}(1+k-\varepsilon)} \,z^{n+k}\,\bar{z}^{k}.
\end{multline}
Let us comment on the calculation of residues.
Assuming the closing of contour in the lower half-plane, we need to
calculate the residues of the function that contains poles of order
$L+1$ at the points $\nu_k = -\iu\frac{n+1}{2} - \iu k, k = 0, 1, \ldots$ (see Remark 2 in section~\ref{sect:2.1} and~\cite[Fig. 10]{DKO}).
The corresponding residue can be expressed as
\begin{align}\label{res}
\mathrm{Res}_{\nu_k} =& \frac{i}{L!}
\left.\partial_{\varepsilon}^{L}\right|_{\varepsilon=0}\,
\left(\frac{\Gamma(1-\varepsilon)\Gamma(1+\varepsilon)}
{\Gamma(\beta+\varepsilon)\Gamma(1-\beta-\varepsilon)}\right)^{L+1}\notag \\
\,&\times\frac{\Gamma^{L+1}(1+n+k-\beta -\varepsilon)\Gamma^{L+1}(1-\beta+k-\varepsilon)}
    {\Gamma^{L+1}(1+k-\varepsilon)\Gamma^{L+1}(1+n+k-\varepsilon)} \,z^{n+k+\frac{1}{2}-\varepsilon}\bar{z}^{\frac{1}{2}+k-\varepsilon}.
\end{align}
The derivation of this formula contains three steps:
\begin{itemize}
\item Calculate integrand at $\nu = \nu_k + \varepsilon$
\begin{align*}
\frac{\Gamma^{L+1}(1+n+k-\beta +\iu\varepsilon)\Gamma^{L+1}(-k-\iu\varepsilon)}
    {\Gamma^{L+1}(\beta-k-\iu\varepsilon)\Gamma^{L+1}(1+n+k+\iu\varepsilon)} \,z^{n+k+\frac{1}{2}+\iu\varepsilon}\bar{z}^{\frac{1}{2}+k+\iu\varepsilon}.
\end{align*}
\item Use the reflection relations for the gamma-function
\begin{align*}
    \Gamma(-k-\iu\varepsilon) =& -\frac{1}{\iu\varepsilon}
\frac{(-1)^k\,\Gamma(1+\iu\varepsilon)\Gamma(1-\iu\varepsilon)}
{\Gamma(1+k+\iu\varepsilon)}; \notag \\
    \Gamma(\beta-k-\iu\varepsilon) =&
\frac{(-1)^{k}\Gamma(\beta - \iu\varepsilon)\Gamma(1 - \beta +\iu\varepsilon)}
{\Gamma(1 - \beta + k + \iu\varepsilon)}\,,
\end{align*}
to transform the previous expression to the form
\begin{align*}
\frac{1}{(-\iu\varepsilon)^{L+1}}
\left(\frac{\Gamma(1+\iu\varepsilon)\Gamma(1-\iu\varepsilon)}
{\Gamma(\beta-\iu\varepsilon)\Gamma(1-\beta+\iu\varepsilon)}\right)^{L+1}\\
\,\frac{\Gamma^{L+1}(1+n+k-\beta +\iu\varepsilon)\Gamma^{L+1}(1-\beta+k+\iu\varepsilon)}
    {\Gamma^{L+1}(1+k+\iu\varepsilon)\Gamma^{L+1}(1+n+k+\iu\varepsilon)} \,z^{n+k+\frac{1}{2}+\iu\varepsilon}\bar{z}^{\frac{1}{2}+k+\iu\varepsilon}
\end{align*}
\item extract the coefficient in front of $\frac{1}{\varepsilon}$
\begin{align*}
\frac{1}{L!}\left.\partial_{\varepsilon}^{L}\right|_{\varepsilon=0}\,
\frac{1}{(-\iu)^{L+1}}
\left(\frac{\Gamma(1+\iu\varepsilon)\Gamma(1-\iu\varepsilon)}
{\Gamma(\beta-\iu\varepsilon)\Gamma(1-\beta+\iu\varepsilon)}\right)^{L+1}\\
\,\frac{\Gamma^{L+1}(1+n+k-\beta +\iu\varepsilon)\Gamma^{L+1}(1-\beta+k+\iu\varepsilon)}
    {\Gamma^{L+1}(1+k+\iu\varepsilon)\Gamma^{L+1}(1+n+k+\iu\varepsilon)} \,z^{n+k+\frac{1}{2}+\iu\varepsilon}\bar{z}^{\frac{1}{2}+k+\iu\varepsilon}
\end{align*}
and the final formula \eqref{res} is obtained after the change $\varepsilon \to \iu \varepsilon$.
\end{itemize}

The Next step is factorization. Using the evident change of the
summation index $p= n+k$ in the first sum, we obtain
\begin{multline}
\sum_{n\in \mathbb{Z}}\sum_{k=0}^{+\infty}
\frac{\Gamma^{L+1}(1-\beta+n+k-\varepsilon)}
{\Gamma^{L+1}(1+n+k-\varepsilon)}
\frac{\Gamma^{L+1}(1-\beta+k-\varepsilon)}
{\Gamma^{L+1}(1+k-\varepsilon)} \,z^{n+k}\,\bar{z}^{k} = \\
\sum_{p\in \mathbb{Z}}
\frac{\Gamma^{L+1}(1-\beta+p-\varepsilon)}
{\Gamma^{L+1}(1+p-\varepsilon)}\,z^{p}\,
\sum_{k=0}^{+\infty}
\frac{\Gamma^{L+1}(1-\beta+k-\varepsilon)}
{\Gamma^{L+1}(1+k-\varepsilon)}\,\bar{z}^{k} = \\
\left[\sum_{p=0}^{\infty}
\frac{\Gamma^{L+1}(1-\beta+p-\varepsilon)}
{\Gamma^{L+1}(1+p-\varepsilon)}\,z^{p} +
\sum_{p=1}^{\infty}
\frac{\Gamma^{L+1}(1-\beta-p-\varepsilon)}
{\Gamma^{L+1}(1-p-\varepsilon)}\,z^{p}\,\right]
\sum_{k=0}^{+\infty}
\frac{\Gamma^{L+1}(1-\beta+k-\varepsilon)}
{\Gamma^{L+1}(1+k-\varepsilon)}\,\bar{z}^{k}
\end{multline}
Note that in the second sum inside the brackets the factor
$\Gamma^{-L-1}(1-p-\varepsilon)$ creates additional $\varepsilon^{L+1}$
so that the whole sum is annihilated by the operator $\left.\partial_{\varepsilon}^{L}\right|_{\varepsilon=0}$ and
after all one obtains
\begin{multline}
\sum_{n \in \mathbb{Z}}\int\limits_{-\infty}^{\infty} d\nu\,\dfrac{\Gamma^{L + 1}(\frac{n+1}{2} - \beta + \iu\nu)\Gamma^{L + 1}( \frac{n+1}{2} - \iu\nu)}{\Gamma^{L + 1}(\frac{n+1}{2} + \beta - \iu\nu)\Gamma^{L + 1}(\frac{n+1}{2} + \iu\nu)}z^{\iu\nu + \frac{n}{2}}\bar{z}^{\iu\nu - \frac{n}{2}}  =
\frac{2\pi}{L!}
\left.\partial_{\varepsilon}^{L}\right|_{\varepsilon=0}\,
(z\bar{z})^{\frac{1}{2}-\varepsilon} \\
\left(\frac{\Gamma(1-\varepsilon)\Gamma(1+\varepsilon)}
{\Gamma(\beta+\varepsilon)\Gamma(1-\beta-\varepsilon)}\right)^{L+1}\,
\sum_{p=0}^{+\infty}
\frac{\Gamma^{L+1}(1-\beta+p-\varepsilon)}
{\Gamma^{L+1}(1+p-\varepsilon)}\,z^{p}\,
\sum_{k=0}^{+\infty}\frac{\Gamma^{L+1}(1-\beta+k-\varepsilon)}
{\Gamma^{L+1}(1+k-\varepsilon)}\,\bar{z}^{k}.
\end{multline}

\section{Derivation of the loop recursion
for the operator $\widetilde{R}_\ell$
in the case $\beta=1$ \label{app:E}}
 First, we introduce instead of
 (\ref{beta0}), (\ref{D=D}) (for $\beta=1$) the function
\be
\lb{RL02b}
\begin{array}{l}
\displaystyle
\widetilde{\widetilde{\Phi}}^{(1)}_L(z,\bar{z}) =
\frac{L!}{\Gamma(\lambda)} (z \bar{z})^{\frac{\lambda-1}{2}} \,
\widetilde{\Phi}^{(1)}_L(z,\bar{z};\lambda) = \\ [0.3cm]
 \displaystyle
= L! \;
\sum\limits_{n=0}^\infty
(n+\lambda)\, C^{(\lambda)}_n
 \left(\dfrac{z+\bar{z}}{2\sqrt{z\bar{z}}}\right)
\int\limits_{-\infty}^{+\infty}
  \dfrac{d\nu \;\; (z\bar{z})^{{\sf i}\nu-1/2}}{
 \Bigl(\frac{1}{4}(\lambda+n)^2-({\sf i}\nu-\frac{1}{2})^2
 \Bigr)^{L+1}} \,,
 \end{array}
\ee
which also depends only on the
conformal variables $z,\bar{z}$. The
operator $(z\partial_z + \bar{z}\partial_{\bar{z}})$
commutes with $\dfrac{z+\bar{z}}{2\sqrt{z\bar{z}}}$
and its action on the function
$\widetilde{\widetilde{\Phi}}_L^{(1)}(z,\bar{z})$ gives
\be
\begin{array}{c}
\displaystyle
(z\partial_z + \bar{z}\partial_{\bar{z}})
\widetilde{\widetilde{\Phi}}_L^{(1)}(z,\bar{z}) =
L! \sum\limits_{n=0}^\infty
(n+\lambda)\, C^{(\lambda)}_n
 \left(z/\bar{z}\right)
\int\limits_{-\infty}^{+\infty}
\dfrac{d\nu \; 2({\sf i}\nu-1/2) \; (z\bar{z})^{{\sf i}\nu-1/2}}{
 \Bigl(\frac{1}{4}(\lambda+n)^2-({\sf i}\nu-\frac{1}{2})^2
 \Bigr)^{L+1}} = \\ [0.3cm]
 = (L-1)!\sum\limits_{n=0}^\infty
(n+\lambda)\, C^{(\lambda)}_n
 \left(z/\bar{z}\right)
\int\limits_{-\infty}^{+\infty}d\nu
\;(z\bar{z})^{{\sf i}\nu-1/2} \;
\partial_{\nu} \; \dfrac{(-i)}{
 \Bigl(\frac{1}{4}(\lambda+n)^2-({\sf i}\nu-\frac{1}{2})^2
 \Bigr)^{L}} = \\ [0.3cm]
 = (L-1)!\sum\limits_{n=0}^\infty
(n+\lambda)\, C^{(\lambda)}_n
 \left(z/\bar{z}\right)
\int\limits_{-\infty}^{+\infty}d\nu \;
 \dfrac{-\log(z\bar{z})
\;(z\bar{z})^{{\sf i}\nu-1/2}}{
 \Bigl(\frac{1}{4}(\lambda+n)^2-({\sf i}\nu-\frac{1}{2})^2
 \Bigr)^{L}} = -\log(z\bar{z})
 \widetilde{\widetilde{\Phi}}_{L-1}^{(1)}
  \; , \\ [0.3cm]
 C^{(\lambda)}_n
 \left(z/\bar{z}\right):=C^{(\lambda)}_n
 \left(\frac{z+\bar{z}}{2\sqrt{z\bar{z}}}\right) \; ,
 \end{array}
\ee
which proves the statement that operator
 (\ref{RL01}) produces the recursion
 $$
 \widetilde{R}_{\ell}
\widetilde{\widetilde{\Phi}}^{(1)}_L(z,\bar{z};\lambda) =
\widetilde{\widetilde{\Phi}}^{(1)}_{L-1}(z,\bar{z};\lambda) \; .
$$

\addcontentsline{toc}{section}{Appendices} 


\renewcommand{\theequation}{\Alph{section}.\arabic{equation}}
\renewcommand{\thetable}{\Alph{table}}
\setcounter{section}{0} \setcounter{table}{0}

\newpage
\bibliography{refs.bib}
\bibliographystyle{JHEP}

\end{document}